\documentclass[aps,twocolumn,nofootinbib,showpacs,final,balancelastpage,superscriptaddress]{revtex4}
\usepackage{graphicx}
\usepackage{tabularx}
\usepackage{amssymb}
\usepackage{amsmath}
\usepackage{amsbsy}
\usepackage{comment}
\usepackage{mathrsfs}
\usepackage{dsfont}
\usepackage{nicefrac}
\usepackage{wrapfig}
\usepackage{amsfonts}
\usepackage{url} 
\usepackage[nolist]{acronym}
\usepackage{nicefrac}
\usepackage{subfigure}
\usepackage{epsf,color,colordvi,pifont}
\usepackage{showkeys}
\DeclareFontFamily{OT1}{pzc}{}
\DeclareFontShape{OT1}{pzc}{m}{it}{<-> s * [1.10] pzcmi7t}{}
\DeclareMathAlphabet{\mathpzc}{OT1}{pzc}{m}{it}
\def\dbar{{\mathchar'26\mkern-12mu \textrm{d}}}

\newcommand{\vecp}{\pmb{p}}
\newcommand{\vecAA}{\pmb{\mathpzc A}}
\newcommand{\tcyc}{\tau_{\mathrm{cycle}}}
\newcommand{\Nplat}{N_{\mathrm{plateau}}}
\newcommand{\donoff}{\delta_{\mathrm{onoff}}}
\newcommand{\ORabi}{\Omega_{\mathrm{Rabi}}}

\newcommand{\boef}{bOEF\space}

\newcommand{\oef}{OEF\space}
\newcommand{\pp}{PP\space}
\newcommand{\odes}{ODEs\space}

\begin{document}

\begin{acronym}[ABCD]
\acro{PP}{pair production}
\acro{bOEF}{bifrequent oscillating electric field}
\acro{ODEs}{Ordinary Differential Equations}
\end{acronym}

\title{Electron-positron pair production in a bifrequent oscillating electric field}
\author{Ibrahim  \surname{Akal}}
\author{Selym \surname{Villalba-Ch\'avez}}
\email{selym@tp1.uni-duesseldorf.de}
\author{Carsten \surname{M\"{u}ller}}
\affiliation{Institut f\"{u}r Theoretische Physik I, Heinrich-Heine-Universit\"{a}t D\"{u}sseldorf, Universit\"{a}tsstra\ss{e} 1, 40225 D\"{u}sseldorf, Germany}

\begin{abstract}
The production of electron-positron pairs from the quantum vacuum polarized by the superposition of a strong and  a perturbative  oscillating  
electric field mode is studied.  Our outcomes rely on a nonequilibrium  quantum field theoretical approach,  described by the quantum kinetic 
Boltzmann-Vlasov equation.  By superimposing the  perturbative mode,  the characteristic resonant effects and Rabi-like frequencies  in the  
single-particle distribution function are modified,  as compared to  the predictions resulting from the case driven   by a strong oscillating  
field mode only. This is demonstrated in the momentum spectra of the produced pairs. Moreover, the dependence of the total number of pairs on 
the intensity parameter of each mode is discussed and a strong enhancement found for large values of the relative Keldysh parameter. 
\end{abstract}



\date{\today}

\maketitle

\section{Introduction}

The spontaneous creation of electron-positron pairs in  a strong external electric field  is  a remarkable nonperturbative phenomenon,  
intrinsically associated with  the  instability of the quantum  vacuum \cite{Sauter:1931,Heisenberg:1935,Schwinger:1951nm}. In a constant electric 
field, this phenomenon--also known as the Schwinger mechanism--has a production  rate  exponentially small 
$\sim\exp\left(-\pi E_c/E\right)$. Hence, its occurrence is expected to be  difficult to achieve experimentally, unless the electric field 
strength $E$ comes close  to  the critical scale  of quantum electrodynamics (QED)   $E_c=1.3 \times 10^{16}\ \rm V/cm$.  Unfortunately, 
a  field of such a  nature  is not within the reach  of  current technical capabilities and   a direct experimental observation of the Schwinger 
mechanism  remains a  big challenge for the contemporary physics.   Hopes  of  reaching the required field strengths in the focal spot of  
envisaged  high-intensity lasers such as the  Extreme Light Infrastructure (ELI)  \cite{ELI} and the Exawatt Center for Extreme Light Studies  
(XCELS) \cite{xcels}  have renewed the  interest in the study of Schwinger-like pair production (PP) processes. Its verification will provide 
significant insights in  the nonlinear QED regime as well as in  various  processes which share its nonperturbative feature. Notably, among them are the  
Unruh and Hawking  radiation and the string breaking  in the theory of the strong interactions \cite{Casher:1978wy}. 

While the prospect of using the strong field of lasers is enticing from practical perspectives, there exists a price to be paid for: the complicated 
nature of the electromagnetic field of a laser pulse--the wave profile, its  space-time dependence  and the existence of magnetic field 
components--introduces   considerable additional complications in the calculations associated with the PP process. Indeed,  it seems 
that a full description of the vacuum decay in such a scenario is far  from  being computationally feasible, at least in the near 
future.  In order to make the problem tractable  some simplifications are required.\footnote{Obviously, the  plane-wave approximation  is excluded, because 
in this case  the field invariants $\mathscr{F}=(E^2-B^2)/2$ and $\mathscr{G}=\pmb{E}\cdot\pmb{B}$ vanish identically, and the vacuum-vacuum transition 
amplitude does not decay against  electron-positron pairs~\cite{Fradkin,Dunne:2004nc}.} In the first  instance, an analysis in a pure electric field  would be  
highly desirable since, on the one hand, it  resembles  the electric-like background used in the genuine Schwinger mechanism, and on the other hand because this  
special field configuration  can  be  obtained to a good approximation through  a  collision of   two  counterpropagating  linearly polarized  laser pulses  with  equal intensities 
and  polarization directions. The resulting field is  a standing  electromagnetic wave  depending separately upon the spatial and temporal coordinates. This setup  substantially 
simplifies  the  treatment of the problem, but the field inhomogenity still represents  a major task to overcome from  analytic and numeric viewpoints.  
Although some progresses in the latter direction have been accomplished already \cite{ruf2009zz,Hebenstreit:2010vz,Hebenstreit:2011wk,Hebenstreit:2013qxa,Berenyi:2013eia}, most of the theoretical efforts carried out 
so far have been focused on the idealization in which the background is a homogeneous electric field oscillating in time \cite{Brezin,Salamin}. As such, this temporal 
dependence renders the  vacuum decay into  electron-positron pairs  a far from equilibrium phenomenon. Consequently, the quantum kinetic theory has turned out 
to be an appropriate formalism for analyzing the aforementioned process \cite{Schmidt:1998vi,Schmidt:1999vi,alkofer:2001ib}.

In parallel,  various mechanisms for compensating the  suppression associated with the spontaneous  creation of electron-positron pairs from  the vacuum have been put 
forward \cite{dgs2008,dgs2009,Muller:2003,Sieczka2006, DiPiazza:2009py,Bulanov:2010gb,Grobe2012,Grobe2013,Titov2013,Jansen2013,Augustin2014,Li2014}. Some of them have been   motivated by the fact  that 
pair production by multiphoton absorption has been observed using nonlinear Compton scattering  \cite{Burke:1997ew}.  In line with this issue, the authors of Refs.~\cite{dgs2008,dgs2009}  
showed  that the pair creation rate can be enhanced by superimposing a weak beam of energetic photons with a strong but low-frequency electric field.  
The study of such a problem  was carried out by taking the slow laser pulse  as a constant electric field reducing the problem to that  of a photon-stimulated 
Schwinger pair creation. The idea has been further extended to the case in which both pulses are Sauter fields with  different time scales 
\cite{Orthaber:2011cm,Fey:2011if}. However, in more realistic setups it is expected that  the subcycle structure of an oscillating electric field  (OEF) 
plays a relevant  role  as  it provides  a phenomenology characterized by resonant effects  and Rabi-like oscillations \cite{popov,Hebenstreit:2009km,Akkermans2012,Narozhny,Avetissian,mocken2010}.  
To the best of our know\-ledge, the study of such a problem in a bifrequent OEF composed of a strong and a weak mode has not been addressed so far. Hence, we aim   
to  show  how these features are manifest in the dynamical-assisted  Schwinger  mechanism and reveal their consequences in the total density of produced pairs. For other 
recent studies of pair production processes in bifrequent electromagnetic fields, we refer the interested reader to Refs.~\cite{Krajewska2012,Wu2014}.

Apart from  this introductory portion,  our  paper has  three additional sections. In  Sec.~\ref{sec:QKA} we describe  some basic features of the PP process  
 within the framework  of  the quantum  Boltzmann-Vlasov equation. Afterwards,  its solution in a  periodic multimode OEF is derived and  a  study
of the resonant effects and Rabi-like oscillations is carried out. The  numerical results are exposed in Sec.~\ref{sec:numerical} for the particular situation 
in which a two-mode OEF  is considered. There, different issues associated with the production mechanism  for various  frequency combinations are  discussed. 
Special attention is paid to the effects caused by the superposition of a strong and perturbative mode on the single-particle distribution function. 
The density of pairs yielded in such a configuration is also investigated. Further comments and remarks  are  finally
given in the Conclusion.

\section{Quantum Kinetic Approach} \label{sec:QKA}

\subsection{The quantum Boltzmann-Vlasov equation \label{subsec:boltzman-vlasov}}

We consider the production of electron-positron pairs from  a vacuum polarized by an external classical electromagnetic field  
which is  described by  a spatially  homogeneous but time-dependent four-potential $\mathpzc{A}_\mu(t)$. With 
the former  assumption   we are implicitly  disregarding  the  potential realization of  avalanche  processes \cite{Fedotov:2010ja} 
that could dim  the  field strength. The latter condition  implies that  the field can be Fourier-expanded in terms 
of  the canonical momentum $\pmb{p}$ and,   additionally,  prevents the existence of  magnetic field components. 
Note that, according to Noether's theorem, the total momentum of each created pair will always sum up to zero. Thus, the creation
of an electron with momentum $\pmb{p}$ is always accompanied by the creation of a positron with momentum $-\pmb{p}$ in the purely
time-dependent external field.

For further convenience, we will 
choose $\mathpzc{A}_\mu(t)$  fulfilling  the temporal gauge--i.e., $\mathpzc{A}_0(t)=0$--so that the nonvanishing electric field is 
given by $\pmb{E}(t)=-d\pmb{\mathpzc{A}}/dt=(0,E(t),0)$ pointing in the $y$ direction. Hereafter, we focus ourselves 
in the subcritical regime $E\ll E_c = m^2/\vert e\vert$,  where $m$ and $e$  
are the electron mass and charge, respectively.\footnote{Here and henceforth we use natural units in which  the  speed of light $c$ 
and the Planck constant  $\hbar$ are equal to unity, $c=\hbar=1$.}  Also,  in  what follows, we  take into account neither the collision 
between the created particles nor their inherent  radiation  fields. Previous investigations 
on these subjects have revealed that  the effects induced by both phenomena  are irrelevant  for the PP whenever  the field strength $E$ 
is weaker than the critical one $E_c$ \cite{Tanji:2008ku,Bloch:1999eu,Vinnik:2001qd}  and this is in fact the regime in which we are 
interested.

We note that  only those Lorentz transformations which leave the external field  invariant   describe the formal invariance  properties of 
the vacuum in the presence of the field.\footnote{This statement is  in line with previous  group-theoretical analyses developed in an  external constant  electromagnetic field~\cite{Bacry1,VillalbaChavez:2012ea} and in the case in which the background is  a circularly polarized plane wave \cite{Richard}.}  Since   they  
form   a   subgroup of the full Lorentz symmetry group and the  concept of one-particle states relies on the irreducible representations of the 
latter group  \cite{Weinberg:1995mt}, one finds that the standard classification of elementary particles  is no longer applicable in the region 
occupied by an external field. Despite this  conceptual  loss,  the  canonical  quantization of  the   matter sector of QED  in  an OEF can be 
carried out \cite{Fradkin}. A relevant step in this direction  results from the  diagonalization of the  corresponding Hamiltonian  in every instant  
of time through time-dependent Bogolyubov transformations. Such a procedure  allows us   to express the spinor field  operator  in terms of degrees 
of freedom  in the external field, i. e., in the so-called quasiparticles representation:
\begin{eqnarray}\label{instantaneousfield}
\begin{array}{c}\displaystyle
\Psi(\pmb{x},t)=\frac{1}{L^{\nicefrac{3}{2}}}\sum_{\pmb{p}} \Phi_{\pmb{p}}(t)e^{i\pmb{p}\cdot\pmb{x}},\\
\displaystyle
\Phi_{\pmb{p}}(t)=\sum_{\mathpzc{s}}\left\{a_{\pmb{p},\mathpzc{s}}(t)\mathpzc{u}_{\pmb{p},\mathpzc{s}}(t)+b_{-\pmb{p},\mathpzc{s}}^\dagger(t)\mathpzc{v}_{-\pmb{p},\mathpzc{s}}(t)\right\},
\end{array} 
\end{eqnarray} where  $ V=L^3$ is  the normalization volume and $\pmb{p}=\frac{2\pi}{L} \pmb{n}$ the  discretized momentum with $\pmb{n}=(n_x,n_y,n_z)$, $n_i=0,\pm1,\pm2,\ldots$  
In this framework,  the time-dependent bispinors 
$\mathpzc{u}_{\pmb{p},\mathpzc{s}}(t)$ and $\mathpzc{v}_{\pmb{p},\mathpzc{s}}(t)$ are eigenfunctions of the boost operator component  along the  external field direction 
with eigenvalues $\mathpzc{s}=\pm\nicefrac{1}{2}$. While $a_{\pmb{p},\mathpzc{s}}(t)$ and  $a_{\pmb{p},\mathpzc{s}}^\dagger(t)$ represent the corresponding 
annihilation and creation operators for  a quasiparticle,  $b_{-\pmb{p},\mathpzc{s}}^\dagger(t)$ and $b_{-\pmb{p},\mathpzc{s}}(t)$ are  the   creation and 
annihilation operators  for an antiquasiparticle, respectively. These two sets of instantaneous second quantization operators satisfy the equal time anticommutation 
relations
\begin{eqnarray}\label{anticommutator1}
&&\left[a_{\pmb{p},\mathpzc{s}}(t),a_{\pmb{p}^\prime,\mathpzc{s}^\prime}^\dagger(t)\right]_+=\delta_{\pmb{p},\pmb{p}^\prime}\delta_{\mathpzc{s},\mathpzc{s}^\prime},\\  
&&\left[b_{\pmb{p},\mathpzc{s}}(t),b_{\pmb{p}^\prime,\mathpzc{s}^\prime}^\dagger(t)\right]_+=\delta_{\pmb{p},\pmb{p}^\prime}\delta_{\mathpzc{s},\mathpzc{s}^\prime}, \label{anticommutator2}
\end{eqnarray}and all other  anticommutators  vanish  identically. Because of their temporal dependences, one can  introduce quantities  that  arise   
naturally in the study of transport phenomena such as the  single-particle distribution function  
\begin{eqnarray}
W(\pmb{p},t)=\sum_{\mathpzc{s}}\langle\mathrm{VAC}, \mathrm{in}\vert a_{\pmb{p},\mathpzc{s}}^\dagger(t)a_{\pmb{p},\mathpzc{s}}(t)\vert\mathrm{VAC}, \mathrm{in}\rangle\label{spdf}
\end{eqnarray} where the ground state $\vert\mathrm{VAC},\mathrm{in}\rangle$  is defined in the Heisenberg picture by 
$a_{\mathrm{in}}\vert\mathrm{VAC},\mathrm{in}\rangle=b_{\mathrm{in}}\vert\mathrm{VAC}, \mathrm{in}\rangle =0$ with  
$a_{\mathrm{in}}\equiv a_{\pmb{p},\mathpzc{s}}$ and $b_{\mathrm{in}}\equiv b_{-\pmb{p},\mathpzc{s}}$ at $t\to t_{\rm in}$. 
The connection between the $\rm in$-operators and the instantaneous ones involved in  Eqs.~(\ref{anticommutator1}) and (\ref{anticommutator2}) 
is mediated by certain  Bogolyubov coefficients $f(\pmb{p},t)$ and $g(\pmb{p},t)$ [see details in Appendix~\ref{subsec:gen-remarks}].  
As a consequence, the time evolution  equations   of  $W(\pmb{p},t)=2\vert f(\pmb{p},t)\vert^2$   follow from the  temporal equations of  
$f(\pmb{p},t)$. The latter  can be  determined  by   exploiting  the fact that the field   representations [Eq.~(\ref{instantaneousfield})] 
satisfy the  Dirac equation in the external field.  This procedure leads to a  system of  coupled  ordinary differential equations (ODEs) 
that have been extensively exploited in the study of  several open questions associated with the \pp process  
\cite{vinnik,Hebenstreit:20091km,Hebenstreit:2009km,mocken2010,Orthaber:2011cm}.  
In particular, by following the notation of Ref.~\cite{mocken2010}, it reads
\begin{eqnarray}\label{firstequa}
&&i\dot{f}(\pmb{p},t)=\mathpzc{a}_{\pmb{p}}(t)f(\pmb{p},t)+\mathpzc{b}_{\pmb{p}}(t)g(\pmb{p},t),\\ 
&&i\dot{g}(\pmb{p},t)=\mathpzc{b}^*_{\pmb{p}}(t)f(\pmb{p},t)-\mathpzc{a}_{\pmb{p}}(t)g(\pmb{p},t)\label{secondequa}
\end{eqnarray}with the initial conditions $f(\pmb{p},-\infty)=0$ and $g(\pmb{p},-\infty)=1$. Note that, hereafter, a dot  indicates a total time derivative. 
The remaining  functions and parameters contained in these formulas  are given by
\begin{eqnarray}\label{coefficient1}
&\displaystyle\mathpzc{a}_{\pmb{p}}(t)=\mathpzc{w}_{\pmb{p}}(t)+\frac{eE(t)p_x}{2\mathpzc{w}_{\pmb{p}}(t)(\mathpzc{w}_{\pmb{p}}(t)+m)},\\
&\displaystyle\mathpzc{b}_{\pmb{p}}(t)=\frac{1}{2}\frac{eE(t)\epsilon_\perp}{\mathpzc{w}_{\pmb{p}}^2(t)}\exp\left[-i\arctan\left(\frac{p_xq_\parallel}{\epsilon_\perp^2+\mathpzc{w}_{\pmb{p}}(t)m}\right)\right],\nonumber\\
\label{coefficient2}
\end{eqnarray}where   the kinetic momentum along $\pmb{E}$  is defined as $q_\parallel(t)=p_\parallel-e\mathpzc{A}(t)$. In this context,  $\epsilon_\perp^2=m^2+\pmb{p}_\perp^{\,2}$ is the transverse
energy squared, whereas  $\mathpzc{w}^2_{\pmb{p}}(t)=\epsilon_\perp^2+q_\parallel^2(t)$ characterizes the total energy squared. Here $\pmb{p}_\perp=(p_x,0,p_z)$ with $p_z=0$ and $\pmb{p}_\parallel=(0,p_y,0)$ are the 
components of the canonical momentum  perpendicular and parallel to the direction of the field, respectively. We emphasize that, due to the cylindrical symmetry of the problem about the $y$ axis,
we may set $p_z=0$ without loss of generality.

Although  Eqs.~(\ref{firstequa})~and~(\ref{secondequa})--with Eqs.~(\ref{coefficient1}) and (\ref{coefficient2}) included--turn  out to be  
appropriate for a numerical assessment, there exists an integrodifferential equation  for $W(\pmb{p},t)$ which allows us to 
extract some important  outcomes from the \pp process:
\begin{eqnarray}\label{vlasov}
&&\dot{W}(\pmb{p},t)=\partial_tW(\pmb{p},t)+eE(t)\partial_{q_\parallel}W(\pmb{p},t)\\&&\qquad\quad=\frac{e E(t)\epsilon_\perp}{\mathpzc{w}_{\pmb{p}}^2(t)}\int_{-\infty}^t dt^\prime \frac{e E(t^\prime)\epsilon_\perp}{\mathpzc{w}_{\pmb{p}}^2(t^\prime)}
[1-W(\pmb{p},t^\prime)]\nonumber\\&&\qquad\quad\times\cos\left[2\int_{t^\prime}^t dt^{\prime\prime}\ \mathpzc{w}_{\pmb{p}}(t^{\prime\prime})\right].\nonumber
\end{eqnarray} The above formula--also  known as the quantum Boltzmann-Vlasov equation--assumes the  vacuum initial condition $W(\pmb{p},-\infty)=0$.
Its  derivation from   Eqs.~(\ref{firstequa})~and~(\ref{secondequa}) is  outlined in Appendix~\ref{appb}. 

Eq.~(\ref{vlasov}) shows  that the \pp is a nonequilibrium time-dependent process.  Besides, it has been recognized that  the combination of  
the nonlocality  in time and the memory effects closely associated with the quantum  statistic factor $\sim 1-W(\pmb{p},t)$ 
provides  Eq.~(\ref{vlasov})  with a pronounced  non-Markovian feature \cite{Schmidt:1998vi,Bloch:1999eu,vinnik}.  It means 
that the single-particle distribution function depends on the number of degrees of freedom  already present in the system. 
We should also mention at this point that the   experimentally  observable  fields are those resulting at asymptotically 
large  times  [$t\to\pm\infty$],  when  the electric field is switched off $\pmb{E}(\pm\infty)\to 0$. These are   
the electron and positron one-particle states to which the degrees of freedom in an  \oef  are relaxed at this asymptotic 
condition. Accordingly,  the asymptotic single-particle  distribution function $W(\pmb{p},\infty)$ is  physically 
meaningful.\footnote{How far the physical interpretation of quasiparticle states in the presence of an external field can 
be stretched has recently been addressed in Ref.~\cite{Dabrowski2014}.}

Finally, we wish to stress that the integration over the momentum, i. e.,
\begin{equation}
\mathpzc{N}_{\ e^-e^+}=\lim_{t\to\infty}\int \dbar^3 p \, W(\pmb{p},t)
\label{eqn:tot-no}
\end{equation}defines the number of produced pairs per unit of volume. Observe that the shorthand notation $\dbar=d/(2\pi)$ has been used here. 
It is worth observing that the corresponding particle creation rate differs from the asymptotic expression of the  vacuum decay rate  per unit 
of volume $\varGamma_{\mathrm{vac}}(t)$  given in Eq.~(\ref{vacdecrate}). Only when  the condition    $\vert f(\pmb{p},\infty)\vert^2\ll1$  is  
fulfilled  one can approach  $\varGamma_{\mathrm{vac}}(\infty)\approx-\dot{\mathpzc{N}}_{\ e^-e^+}$. However, we will see very shortly that--owing 
to resonant effects--the former limit is not always satisfied in an OEF, 
requiring a clear distinction  between both concepts.

\subsection{Resonance effects and Rabi-like oscillations in a multimode standing electric wave} \label{subsec:res-effects}

The  PP  in an OEF   is  characterized   by resonance effects associated with the absorption of  
multiple energy packages [``photons''] from the external field  and  by  Rabi-like oscillations. This result has been found by Popov \cite{popov} and 
further developed by other authors \cite{Narozhny, Avetissian, mocken2010,BlaschkeCPP} for a single-mode OEF. In this subsection, we show how both properties  are 
manifest in the presence  of a multimode OEF and within the framework of nonequilibrium quantum field theory.  To this end, we will study an equivalent 
representation of Eq.~(\ref{vlasov}) [see details in  Appendix~\ref{appb}]: 
\begin{eqnarray}\label{intermdiate1v}
&\displaystyle\dot{\bar{f}}(\pmb{p},t)=-\frac{eE(t)\epsilon_\perp}{2\mathpzc{w}_{\pmb{p}}^2(t)}\bar{g}(\pmb{p},t)\exp\left[2i\int_{t_0}^t dt^\prime\ \mathpzc{w}_{\pmb{p}}(t^\prime)\right],\\ 
&\displaystyle\dot{\bar{g}}(\pmb{p},t)=\frac{eE(t)\epsilon_\perp}{2\mathpzc{w}_{\pmb{p}}^2(t)}\bar{f}(\pmb{p},t)\exp\left[-2i\int_{t_0}^t dt^\prime\ \mathpzc{w}_{\pmb{p}}(t^\prime)\right],
\label{intermdiate2v}
\end{eqnarray}in which the lower integration limit $t_0$ sets an arbitrary phase at a given instant of time and $W(\pmb{p},t)=2\vert \bar{f}(\pmb{p},t)\vert^2$. Next,  we  decompose the vector 
potential comprising $k$ modes according to
\begin{equation}
\mathpzc{A}_\mu(\eta_1,\ldots,\eta_k)=\sum_{i=1}^{k}\mathpzc{A}_\mu^{(i)}(\eta_i),\quad\mathrm{with}\quad \eta_i=\omega_it.
\end{equation} Each mode $\mathpzc{A}_\mu^{(i)}(\eta_i)$   is supposed to be a $2\pi-$periodic function in the variable $\eta_i$. 
The resulting  field  $\mathpzc{A}_\mu(\eta_1,\ldots,\eta_k)$  does not have  a well-defined periodicity in time. However,   it turns out to 
be a periodic function in each variable $\eta_1$, $\eta_2,\dots$, $\eta_k$  separately. By taking advantage  of this fact,  a periodic part  
$\tilde{\Theta}_{\pmb{p}}(\eta_1,\ldots,\eta_k)$ can be separated in the dynamical phase 
$\int_{t_0}^t dt^{\prime}\ \mathpzc{w}_{\pmb{p}}(t^{\prime})=\bar{\varepsilon}_{\pmb{p}}t+\tilde{\Theta}_{\pmb{p}}(\eta_1,\dots,\eta_k)$, 
with $\bar{\varepsilon}_{\pmb{p}}$ being the  electron quasienergy. In correspondence, we  expand the product of functions contained in  
Eqs.~(\ref{intermdiate1v}) and (\ref{intermdiate2v}) in Fourier series:  
\begin{eqnarray}\label{fourierexpansion}
&& \frac{eE(t)\epsilon_\perp}{\mathpzc{w}_{\pmb{p}}^2(t)}\exp\left[2i\int_{t_0}^t dt^{\prime}\ \mathpzc{w}_{\pmb{p}}(t^{\prime})\right]\simeq \sum_{\mathpzc{n}_1\ldots\mathpzc{n}_k=-\infty}^\infty \Lambda_{\mathpzc{n}_1,\ldots,\mathpzc{n}_k}(\pmb{p})\nonumber\\ 
&& \qquad\qquad\qquad\times \exp\left[2i\bar{\varepsilon}_{\pmb{p}}t-i\sum_{j=1}^k \mathpzc{n}_j \eta_j \right].
\end{eqnarray} Here the Fourier coefficients are $k$-fold parametric integrals given by
\begin{eqnarray}
&&\Lambda_{\mathpzc{n}_1,\ldots,\mathpzc{n}_k}(\pmb{p})=\int_{-\pi}^{\pi}\dbar\eta_1\ldots\int_{-\pi}^{\pi}\dbar\eta_k\frac{eE(\eta_1,\ldots,\eta_k)\epsilon_\perp}{\mathpzc{w}_{\pmb{p}}^2(\eta_1,\ldots,\eta_k)}\nonumber\\ 
&&\qquad\qquad\times\exp\left[2i\tilde{\Theta}_{\pmb{p}}(\eta_1,\ldots,\eta_k)+i\sum_{j=1}^k\mathpzc{n}_j\eta_j\right]
\end{eqnarray} whose explicit  expression is not important to show the generic nature of the process.

Note that the only place where the time enters in Eq.~(\ref{fourierexpansion}) is  in the exponentials. Because of this, they will 
oscillate wildly as the limit $t\to\pm\infty$ is taken into account. As such, the combination  which  minimizes  the exponent  
is dominant and   promotes the energy conservation  in the \pp processes
\begin{equation}
2\bar{\varepsilon}_{\pmb{p}}=\sum_{j=1}^k n_j\omega_j.
\label{eqn:resonance-condition-gen}
\end{equation}
The corresponding Fourier indices are denoted by $n_1,\ldots,n_k$ here.
When fast-varying  terms are dropped,  Eq.~(\ref{fourierexpansion}) can be approached by its most slowly varying Fourier 
mode\footnote{For commensurable field frequencies $\omega_1,\ldots,\omega_k$, there 
can be more than one exact solution of Eq.~(\ref{eqn:resonance-condition-gen}). Also, in the 
incommensurable case, several integer combinations of frequencies may 
solve Eq.~(\ref{eqn:resonance-condition-gen}) approximately (see Sec.III-C). Hence, in general, there can be more 
than just one dominant Fourier mode.}
\begin{eqnarray}
&&\frac{eE(t)\epsilon_\perp}{\mathpzc{w}_{\pmb{p}}^2(t)}\exp\left[2i\int_{t_0}^t dt^\prime\ \mathpzc{w}_{\pmb{p}}(t^\prime)\right]\approx  \Lambda_{n_1,\ldots,n_k}(\pmb{p})\nonumber\\&&\quad\qquad\qquad\qquad\qquad\qquad\times \exp\left[i\Delta_{n_1,\ldots,n_k}(\pmb{p})t\right]\label{fourierexpansionapproach}
\end{eqnarray}with  $\Delta_{n_1,\ldots,n_k}(\pmb{p})\equiv2\bar{\varepsilon}_{\pmb{p}}-\sum_jn_j\omega_j$ being  the detuning parameter. Owing to this approximation, Eqs.~(\ref{intermdiate1v}) 
and (\ref{intermdiate2v}) reduce to  an  ODE system  whose solutions can be found  without too much effort. Indeed, the resulting  $\bar{f}(\pmb{p},t)$ allows 
us to express the single-particle distribution function as
\begin{eqnarray}
&& W_{n_1,\ldots,n_k}(\pmb{p},t)\approx\frac{1}{2}\frac{\vert\Lambda_{n_1,\ldots,n_k}(\pmb{p})\vert^2}{\Omega_{\mathrm{Rabi}}^2(\pmb{p})}\nonumber\\ 
&&\qquad\qquad\qquad\qquad\times\sin^2\left[\Omega_{\mathrm{Rabi}}(\pmb{p})(t-t_\mathrm{in})\right]\label{resonantdistributionfunction}\ ,\nonumber
\end{eqnarray}
where  we have supposed  that the field is suddenly turned on at $t_\mathrm{in}$ with $\bar{f}(\pmb{p},t_\mathrm{in})=0$ and $\bar{g}(\pmb{p},t_\mathrm{in})=1$. 
Here the  Rabi-like frequency of the vacuum is given by
\begin{equation}\label{rabifrequency}
 \Omega_{\mathrm{Rabi}}(\pmb{p})\equiv\frac{1}{2}\left[\vert\Lambda_{n_1,\ldots,n_k}(\pmb{p})\vert^2+\Delta_{n_1,\ldots,n_k}^2(\pmb{p})\right]^{\nicefrac{1}{2}}.
\end{equation} 

That  the single-particle distribution function oscillates with this frequency  is a  clear manifestation of the vacuum instability 
in a multimode OEF.  This statement can be verified by supposing that the standing wave is instantaneously turned 
off after the  interaction time $\tau=t_\mathrm{out} - t_\mathrm{in}$. For simplicity we set  the   momenta to zero $\pmb{p}=0$  
and study Eq.~(\ref{resonantdistributionfunction}) near resonance $\Delta_{n_1,\ldots,n_k}\simeq0$.\footnote{Away from the resonance, the oscillations are faster, but their amplitude is lower.} In such a case, 
the  Rabi-like frequency of the vacuum approaches  $\Omega_{\mathrm{Rabi}}^{(0)}\equiv\Omega_{\mathrm{Rabi}}(0)\approx \frac{1}{2} \vert\Lambda_{n_1,\ldots,n_k}(0)\vert$ 
and the distribution function  acquires the form  
\begin{eqnarray}\label{rabioscillationvacuumquasiparticle}
W_{n_1,\ldots,n_k}(t)\approx \left\{\begin{array}{cc}
2\sin^2
\left[\Omega_{\mathrm{Rabi}}^{(0)}(t-t_\mathrm{in})\right],\ & t<t_\mathrm{out}\\ \\ 
2\sin^2\left[\Omega_{\mathrm{Rabi}}^{(0)}\tau\right],\ & t\geqslant t_\mathrm{out}
\end{array}\right.\ .
\end{eqnarray}

The above  formula  shows  an oscillatory pattern resulting from  continuous  transitions characterized  by  a period 
$\mathpzc{T}=2\pi/\Omega_{\mathrm{Rabi}}^{(0)}$. This effect resembles the  Rabi oscillation associated 
with a driven two-level atomic system.  In accordance,   Eq.~(\ref{rabioscillationvacuumquasiparticle}) provides  an  evidence   
that in the  field  of a multimode OEF,  the number of quasiparticles  in a  vanishing momentum state is not  stationary.   
Clearly,  for times larger than the interaction time [$t>\tau$], the distribution function for the asymptotic 
states  is constant in time, which indicates  that both the quantum  vacuum and the created electron-positron pairs 
reach the required stability to carry out experimental measurements.

\section{Numerical aspects and results}
\label{sec:numerical}
In this section we perform a detailed numerical analysis of the \pp process in a bifrequent oscillating electric field (bOEF).
To this end, we have implemented a C++ code capable of solving the system of coupled \odes given in Eqs.~\eqref{firstequa} 
and \eqref{secondequa}. Our goal is to investigate the assisted Schwinger mechanism resulting from the combination of a strong 
and a perturbative field mode. To make a clear distinction between these two different modes it is convenient to introduce the 
dimensionless intensity parameters
\begin{equation}
 \xi_j = \frac{e E_j}{m \omega_j}\quad\mathrm{with}\quad j \in \{1,2\}.
\end{equation}
Here $\omega_j$ and $E_j$ are the frequency and the electric field amplitude, respectively, of the $j$th mode. For the strong mode, 
we shall set the field attributes to $\xi_1 = 1.0$ and $\omega_1 = 0.3m$,  which  leads to a field  strength of $E_1=0.3 E_c$.  In contrast, for the perturbative mode,
we shall apply  $\xi_2 \ll 1$ and a variable frequency $\omega_2$.

Let us put the chosen field parameters into perspective. Today,  a strong wave with
the described characteristics is  beyond the reach of  the existing laser technology,
even for the ELI and XCELS projects \cite{ELI}, where $E\lesssim 10^{-2} E_c$
in the optical regime $\omega\sim 1\ \rm eV$ is envisaged corresponding to
$\xi\lesssim 10^3$. Substantially higher photon energies of the order of
$\omega\lesssim 10 \ \rm keV$ are available at modern x-ray free-electron laser
facilities such as the Linac Coherent Light Source at SLAC (Stanford, California).
But the maximum field strengths there lie 3 to 4 orders of magnitude below $ E_c$,
corresponding to $\xi\lesssim 10^{-2}$ (see Sec. II. in Ref.~\cite{DiPiazza:2011tq}
and references therein). The motivation for our choice of parameters is, on the
one hand, to allow for better insights into the PP process in a boEF by highlighting its
intrinsic phenomenology and, on the other hand, to render the numerical
computations feasible.

In order to avoid numerical inaccuracies at starting and ending points of the bOEF, we choose a modulated linear polarized potential of the form
\begin{align}
 \pmb{\mathpzc{A}}(\eta_1) = &- \frac{m \xi_1}{e} \sin(\eta_1) F_1(\eta_1) \hat{\pmb{y}}\nonumber \\ 
 &- \frac{m \xi_2}{e} \sin(\Delta\omega \eta_1) F_2(\eta_1) \hat{\pmb{y}},
 \label{eqn:gauge-field}
\end{align}
where $\eta_1 = \omega_1 t$, $\Delta \omega = \omega_2 / \omega_1$ and $\hat{\pmb{y}} = (0,1,0)^{T}$ defines the polarization direction 
of the field. In Eq.~\eqref{eqn:gauge-field} the envelope functions $F_1(\eta_1)$ and $F_2(\eta_1)$ generally allow us to construct a 
field of any specific time duration separately for both modes. In the present study, however, we shall assume a uniform envelope function 
$F_1(\eta) = F_2(\eta) \equiv F(\eta)$, with $\sin^2$-shaped turn-on and turn-off segments and a plateau region of constant field intensity 
in between. It is given by
\begin{equation}
 F(\eta) = \begin{cases}\sin^2\left(\frac{\eta}{4 \donoff}\right) & 0 \leqslant \eta < 2 \pi \donoff\\ 
1 & 2 \pi \donoff \leqslant \eta \leqslant 2 \pi K\\ 
\sin^2\left(\frac{N\pi}{2\donoff} - \frac{\eta}{4\donoff}\right) & 2 \pi K < \eta \leqslant 2 \pi N\\ 
0 & \mbox{otherwise} \end{cases} ,
\label{eqn:envelope}
\end{equation}
where $N = \Nplat + 2 \donoff$ and $K = N - \donoff$ hold. The lengths of the turn-on and turn-off segments of the field will be chosen 
throughout as $\donoff = 0.5$. The plateau region comprises $\Nplat$ field cycles. Consequently, the resulting total time duration of the 
\boef is given by $N = \Nplat + 1$  in units of the strong mode period $\tcyc = 2 \pi / \omega_1$. 

Notice, that Eq.~\eqref{eqn:gauge-field} with Eq.~\eqref{eqn:envelope} included guarantees the starting of the bOEF with zero amplitude at 
$t=0$. Besides, in order to improve the stability of the numerical integration scheme, all computations have been started half a period 
earlier than $t=0$ when there is no field present. Similarly, all computations have been ended half a period after the external field has 
been switched off.

\subsection{Resonance spectrum}
\label{subsec:res-spec}

We start our analysis by studying the dependence of the single-particle distribution function on the frequency of the weak mode. 
To this end, the latter has been varied within the interval $0.2m\le\omega_2\le 2.5m$. Here, we first consider the case that the 
particles are created with zero momentum ($\pmb{p}=0$). The particles' momentum spectra will be discussed in Sec.III.C below.

\begin{figure}[!h]
 \centering
 \rule{0.48\textwidth}{0.5pt}\\[1.0em]
 \includegraphics[width=0.45\textwidth]{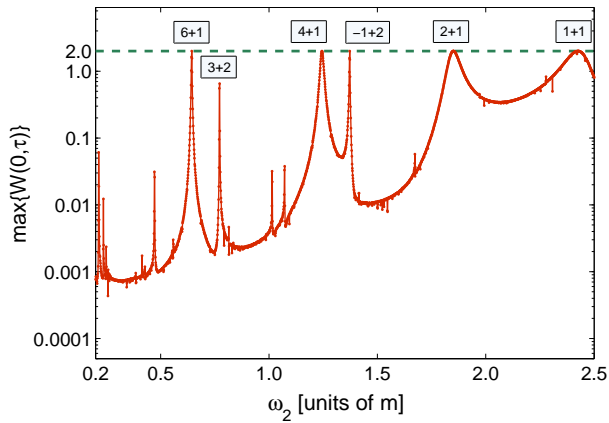}
 \caption{Resonance spectrum for pair production in a bOEF. Shown are the maximum values of $W(0,\tau)$ (red line) for varying plateau 
 lengths from $\Nplat=0$ to 100 versus the perturbative mode frequency of the field. The maximum possible value of the single-particle 
 distribution function is indicated by the dashed horizontal green line. The remaining field parameters are $\xi_1 = 1.0$, $\xi_2 = 0.1$ 
 and $\omega_1 = 0.3m$. Note the logarithmic scale of the ordinate.}
 \label{fig:resonances}
 \rule{0.48\textwidth}{0.5pt}
\end{figure}

For a given value of $\omega_2$, the number of plateau cycles $\Nplat$ has been varied from 0 to 100 in one-cycle steps and the maximum 
value of $W(0,\tau)$, with $\tau=N\tcyc$, has been recorded. This way, we are not sensitive to the Rabi-like time dependence of the 
single-particle distribution function, but rather focus on its maximally achievable values, depending on the secondary mode frequency 
$\omega_2$.  The outcome of this setting possesses a very clear resonant behavior [see Fig.~\ref{fig:resonances}]. At some specific positions 
there are peaks achieving maximal values [$W(0,\tau) = 2.0$] after the interaction time $\tau$, whereas less pronounced resonances also occur. 

We wish to identify and characterize the resonances in terms of the number of photons,  which are absorbed from the strong and perturbative 
modes. To this end we compute the effective mass $m_*=\lim_{\pmb{p}\to0} \bar \varepsilon_{\pmb{p}}$ defined from the quasi-energy
\begin{equation}
\bar \varepsilon_{\pmb{p}} = \frac{1}{2 \pi N_{\mathrm{max}}} \int_{0}^{2 \pi N_{\mathrm{max}}} \sqrt{m^2 + [\pmb{p} - e \vecAA (\eta_1)]^2}\ d\eta_1 .
\label{eqn:quasi-energy}
\end{equation}
Here, $N_{\mathrm{max}}$ denotes the plateau length of the \boef for which the maximal value of $W(0,\tau) \approx 2.0$ is achieved for 
the first time. It turns out that resonances occur whenever the relation 
\begin{equation}
 2 m_* \approx n_1 \omega_1 + n_2 \omega_2
\label{eqn:resonance-condition}
\end{equation}
is fulfilled, which corresponds to the energy conservation law in Eq.~\eqref{eqn:resonance-condition-gen} for $\pmb{p}=0$.
This is illustrated in Table \ref{tab:resonances}. 

We find that the strong field mode dominates the pair-production process in the sense that, in most cases, it contributes more photons 
than the weak mode. This is because it is easier to absorb photons from the strong than from the weak mode since $\xi_1 \gg \xi_2$. The 
perturbative mode typically provides just a single photon during the resonant process. Moreover one can observe that $n_1$ rises in double 
steps from [2+1] to [4+1] and [6+1] while $\omega_2$ decreases by roughly $\Delta\omega_2 \approx 0.6m$. This coincides with the fact, 
that these two additional photons, with $\omega_1 = 0.3m$ each, compensate the required energy difference, which confirms the classification 
scheme based on Eq.~\eqref{eqn:resonance-condition}. Besides, Table I shows that $N_{\mathrm{max}}$ is getting substantially larger for 
decreasing $\omega_2$ along the series of resonances [2+1], [4+1], and [6+1]. 

It is interesting to note that  resonances with $n_2=2$ also occur. Since the weak mode enters in second order into these processes, the 
corresponding peaks are very narrow and the Rabi period becomes long. The latter is evident by the large value of $N_{\rm max}$ for the [3+2] 
 resonance, which we could determine only by substantially  increasing the interacting time, and  by a comparison of $N_{\rm max}$ for the $3-$photon 
 resonances, i. e. [2+1] vs [-1+2]. When the energy $\omega_2$ of the photons from the weak mode is large enough, it even becomes possible 
that pairs are produced by the absorption of two of them with simultaneous emission of a low-frequency photon from the strong mode. This is 
exemplified by the resonance [-1+2]. Further peaks, which are not labeled in Fig.~\ref{fig:resonances}, do not have the strength to reach 
the maximum value of $W(0,\tau)\approx 2$ during the range of interaction times under consideration which is restricted by $N_{\mathrm{max}} \le 100$.

\begin{table}[h]
\rule{0.48\textwidth}{0.5pt}
\caption{Number of photons $n=n_1 + n_2$ versus the perturbative mode frequency $\omega_2$. $N_{\mathrm{max}}$ is the plateau length of the 
\boef for which the maximal value of $W(0,\tau)$ is achieved for the first time.}
\begin{center}
\begin{tabular}{|p{1.7cm}|p{1.7cm}|p{1.7cm}|p{1.7cm}|}
\hline
\parbox[0pt][1.8em][c]{0cm}{} $n=n_1 + n_2$ & $\omega_2\ [m]$ & $N_{\mathrm{max}}$ & $2 m_*\ [m]$\\
\hline
\hline
\ 6+1 & 0.64293 & 76 & 2.4311\\
\hline
\ 4+1 & 1.24385 & 16 & 2.4181\\
\hline
\ 2+1 & 1.84968 & 4 & 2.3775\\
\hline
\ 3+2 & 0.77160 & 212 & 2.4337\\
\hline
-1+2 & 1.37200 & 43 & 2.4285\\
\hline
\hline
\end{tabular}
\end{center}
\label{tab:resonances}
\rule{0.48\textwidth}{0.5pt}
\end{table}

According to Fig.~\ref{fig:resonances} and Table~\ref{tab:resonances}, solely odd total photon numbers $n= n_1 + n_2$ occur in general. An 
explanation for this feature can be obtained from the symmetry of charge conjugation (C parity) for real photons. Based on the setting 
$\vecp = 0$ one can show that the Cparity of the produced pair should be odd \cite{mocken2010,c-parity}. This property can only be realized, 
if the total number of absorbed photons resembles the same feature, since the charge conjugation of a single photon is also odd. Thus, this 
real-photon property coincides very well with our photon picture. One has to note, however, that this reasoning does not cover all cases, as 
the appearance of a [1+1] resonance indicates (see also Fig.~2 in Ref.~\cite{mocken2010}). Due to its large width, this resonance is not very well 
defined, though. Therefore we refrain from further discussion of this particular case.

\subsection{Rabi-like oscillations}
\label{subsec:rabi-like}

\begin{figure}[!h]
 \centering
 \rule{0.48\textwidth}{0.5pt}
 \includegraphics[width=0.51\textwidth]{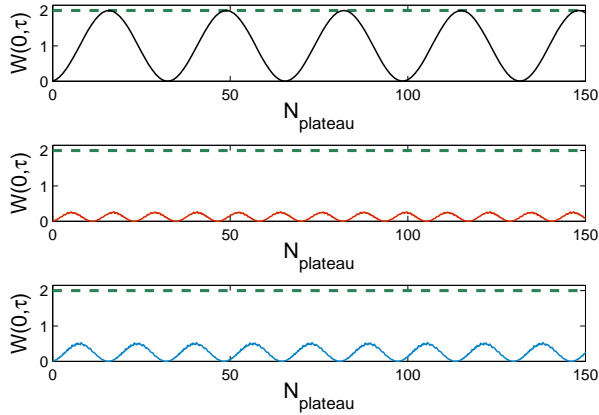}
 \caption{Resonant Rabi-like oscillation in comparison with off-resonant \pp process: The first curve (top to bottom) shows the single-particle 
 distribution function $W(0,\tau)$ versus the plateau length $\Nplat$ for the [4+1]-process. The same parameters as in Fig.~\ref{fig:resonances} 
 with $\omega_2 = 1.24385m$ are used. The second and third graphs refer to $\omega_2 = 1.22m$ and $\omega_2 = 1.26m$, respectively, as the 
 perturbative mode frequency of the \boef.}
 \label{fig:rabis}
\rule{0.48\textwidth}{0.5pt}
\end{figure}

As shown in Sec.~II~B, certain frequency combinations exhibit Rabi oscillations with maximal amplitude in the form of $2 \sin^2(\ORabi \tau)$ 
[see Eq.~\eqref{rabioscillationvacuumquasiparticle}]. This behavior is shown in Fig.~\ref{fig:rabis} for the resonance [4+1] from Table~I. 
To highlight the difference from off-resonant PP processes, we present two more graphs for the nearby frequencies $\omega_2 = 1.22m$ and 
$\omega_2 = 1.26m$, which lie slightly below and  above the resonance, respectively.

From the top curve in Fig.~\ref{fig:rabis} we can extract the corresponding Rabi-like frequency by fitting the $\sin^2$ function from above to the 
underlying numerical data. In this context, $\ORabi$ is the fit parameter to be determined. Such a procedure, carried out for other resonances 
as well, leads to the set of values shown in Table~\ref{tab:ORabis}. Since the Rabi-like frequency is inversely proportional to $N_{\mathrm{max}}$, 
we find that $\ORabi$ decreases along the series of resonances [2+1], [4+1], and [6+1]. Besides, the results of our fitting procedure confirm 
the relation $\ORabi \ll \omega_{1,2}$, which is required for the validity of the resonance condition \cite{mocken2010}. \\

Resonances also exist for nonzero particle momenta. They can be characterized by the same method, with only the effective mass $m_*$ in 
Eq.~\eqref{eqn:resonance-condition} being replaced by the quasienergy $\bar\varepsilon_{\pmb{p}}$ of Eq.~\eqref{eqn:quasi-energy}. Our 
corresponding results will be discussed in the following subsection.

\subsection{Momentum distributions in longitudinal and transversal direction}
\label{subsec:1d-mom}

Further insights into the PP process in a bOEF can be gained from a consideration of the momentum distributions of the created particles. First 
we shall examine one-dimensional momentum spectra and distinguish between the longitudinal momentum component $p_y$ (along the field) and the 
transversal momentum component $p_x$. Recall that an electron of momentum $\pmb{p}$ is always created jointly with a positron of momentum $-\pmb{p}$ 
and that we have set $p_z=0$ without loss of generality. We perform our analysis for the field parameters which lead to the [4+1] resonance when 
the particles have zero momentum. In particular, this means $\omega_2 = 1.24385m$ and $\Nplat = 16$. Equations~\eqref{firstequa} and \eqref{secondequa} 
have been solved for momentum values of $p_x$ and  $p_y$, respectively, varying from $0$ to $2.5m$ in steps of $\Delta p = 0.01 m$.

\begin{table}[!h]
\rule{0.48\textwidth}{0.5pt}
\caption{Rabi-like frequencies for various PP resonances, expressed in units of the fixed frequency $\omega_1 = 0.3m$ of the strong mode and the 
varying frequency $\omega_2$ of the weak mode, respectively.}
\begin{center}
\begin{tabular}{|p{1.7cm}|p{1.7cm}|p{1.7cm}|p{1.7cm}|}
\hline
\parbox[0pt][1.8em][c]{0cm}{} $n=n_1 + n_2$ & $\omega_2\ [m]$ & $\ORabi / \omega_1$ & $\ORabi / \omega_2$\\
\hline
\hline
\ 6+1 & 0.64293 & 0.0033 & 0.0015\\
\hline
\ 4+1 & 1.24385 & 0.0153 & 0.0037\\
\hline
\ 2+1 & 1.84968 & 0.0579 & 0.0094\\
\hline
\ 3+2 & 0.77160 & 0.0012 & 0.0005\\
\hline
-1+2 & 1.37200 & 0.0056 & 0.0012\\
\hline
\hline
\end{tabular}
\end{center}
\label{tab:ORabis}
\rule{0.48\textwidth}{0.5pt}
\end{table}

\begin{figure}[!h]
\rule{0.48\textwidth}{0.5pt}\\
\subfigure[\ Momentum distribution in transversal direction]{\includegraphics[width=0.48\textwidth]{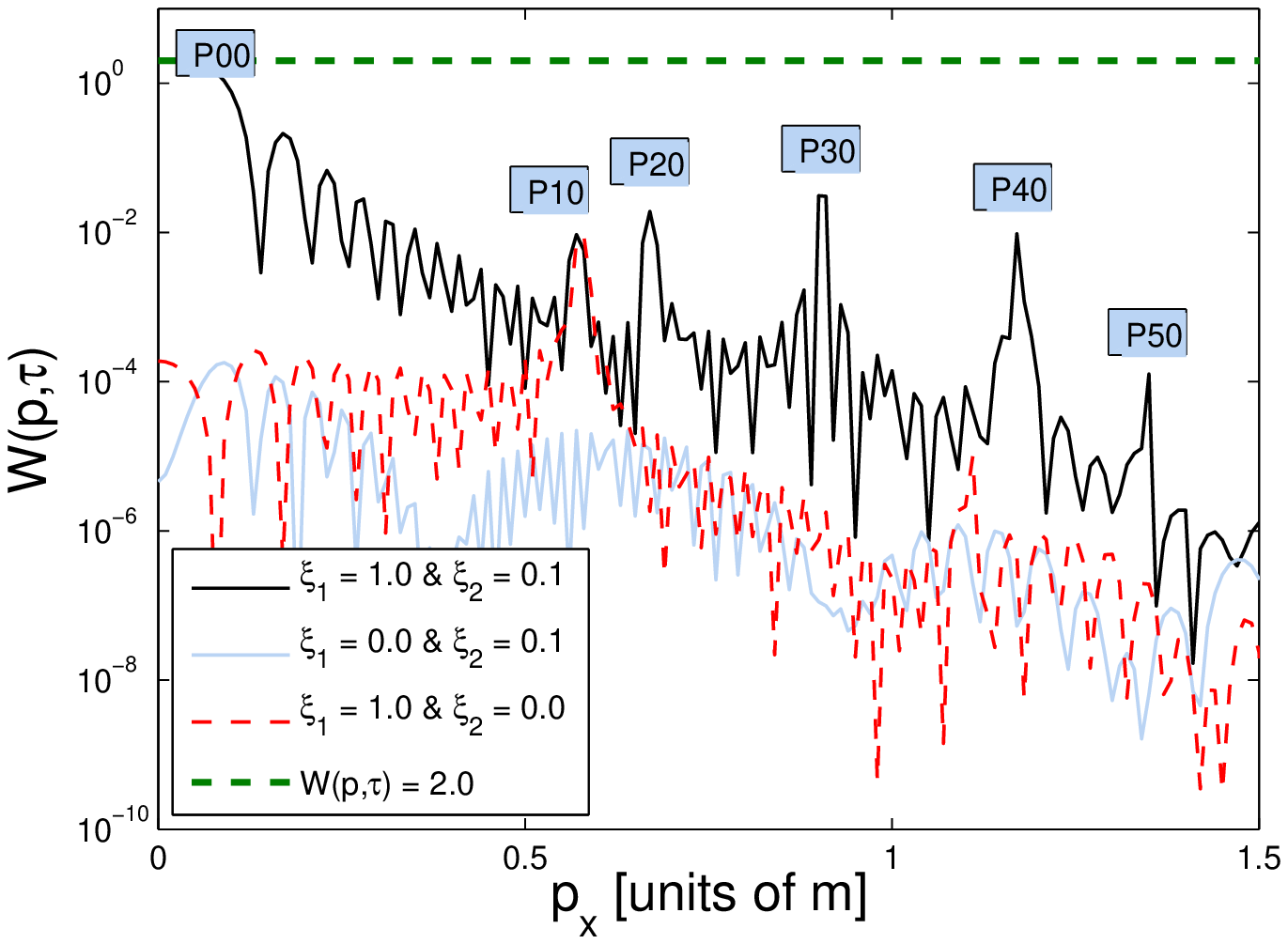}}\hfill
\subfigure[\ Momentum distribution in longitudinal direction]{\includegraphics[width=0.48\textwidth]{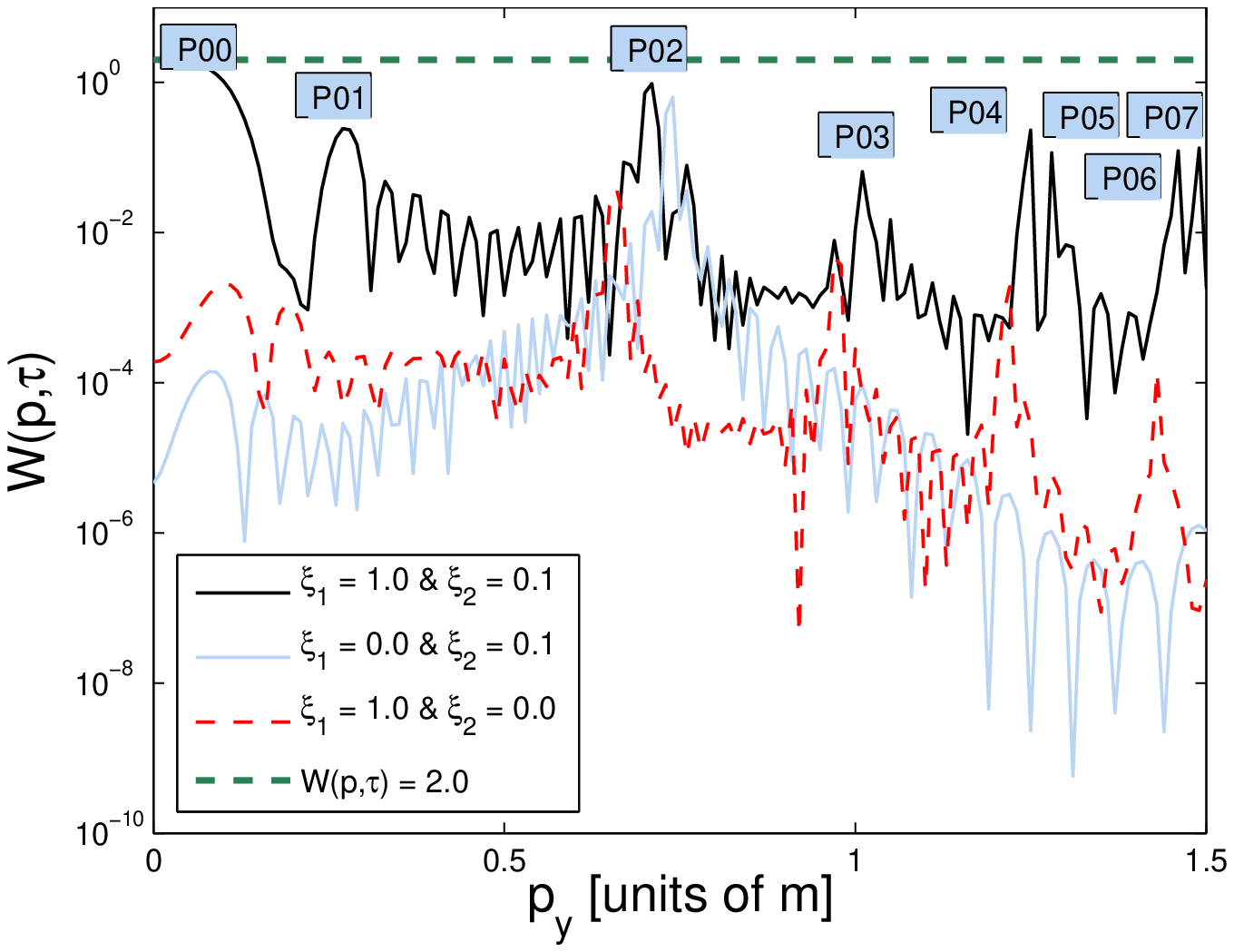}}
\caption{One-dimensional momentum distributions of particles created in a bOEF with $\xi_1 = 1.0$, $\xi_2 = 0.1$, $\omega_1 = 0.3m$, $\omega_2 = 1.24385m$, 
and plateau length $\Nplat = 16$ (black solid lines). The single-particle distribution function is plotted on a logarithmic scale versus (a) the transversal 
momentum $p_x$ and (b) the longitudinal momentum $p_y$. Resonance peaks are labeled with the relevant photon numbers. For comparison, corresponding momentum 
distributions of particles produced in mOEFs are also shown where we have set $\xi_1=0$ [blue (grey) solid lines] and $\xi_2=0$ (red dashed lines), respectively.}
\label{fig:1dim-mom}
\rule{0.48\textwidth}{0.5pt}
\end{figure}

Our results are shown by the black solid lines in Figs.~\ref{fig:1dim-mom}(a)~ and~\ref{fig:1dim-mom}(b). We notice that for the chosen set of field 
parameters both curves start at the maximum $W(\pmb{p}=0,\tau)=2$. For increasing momenta, the single-particle distribution function generally attains 
smaller values, whereby for some specific momenta the curves exhibit resonance peaks. By inspection of the corresponding quasienergies one can 
associate specific numbers of photons with these resonant structures. For the most pronounced peaks, they are compiled in Table~\ref{tab:photonNoVsMomenta}. 
We note that even total photon numbers now also occur in several cases. The reason is that the previous C-parity selection rule is abrogated by the 
fact that nonvanishing momenta may lead to nonvanishing orbital angular momenta $\ell \neq 0$. Hence, the C parity of a produced pair may be even.

Interestingly,  for the appearance of a peak, both modes are not always responsible together. This is exemplified by the resonance P10 in Fig.~\ref{fig:1dim-mom}(a), 
whose energy $2\bar\varepsilon_{\pmb{p}}$ indicates that it originates from the absorption of nine photons from the strong mode and no photon from the weak mode. This 
characterization is confirmed by the appearance of the same resonance peak when the weak mode is switched off (red dashed line). Furthermore, we note that some of 
the observed resonance energies cannot be related uniquely to a specific combination of frequencies. Since the energy of four photons from the strong mode lies close 
to the energy of a single photon from the weak mode, several peaks are likely to consist of two closely spaced resonances, which are not resolved by our calculations. 
In particular, the resonances P40 and P50 possess shoulders at the left side of the main peak which may be attributed to the contribution from a second frequency 
combination. While this phenomenon also occurs in the longitudinal momentum distribution in Fig.~\ref{fig:1dim-mom}(b), here the corresponding pairs of resonances 
at high momenta are resolvable into separate peaks (see P04-P07). We point out that the red dashed line in Fig.~\ref{fig:1dim-mom}(b) nicely shows the monofrequent 
resonances with $n=9,10,11$, and 12 in the strong OEF mode alone. The blue (grey) solid line also shows a very pronounced resonance (close to the P02 peak) for the 
case when only the weak mode drives the vacuum decay. It corresponds to the absorption of two high-frequency photons; note that, in contrast to the nearby P02 peak, 
here the resonance energy at $p_y\approx 0.7m$ amounts to $2\bar\varepsilon_{\pmb{p}}\approx 2.44m$ only because the strong mode does not contribute to the field 
dressing of the energy in Eq.~(\ref{eqn:quasi-energy}).

Comparing Fig.~\ref{fig:1dim-mom}(a) with Fig.~\ref{fig:1dim-mom}(b), one observes that large momenta along the transverse direction are, as a general trend, more 
strongly suppressed than large longitudinal momenta. This can be understood by noting that the latter enter into the quasienergy through the term $p_y-e\mathpzc{A}$, so that 
large values of $p_y$ are partially compensated by the field. Hence, due to the symmetry of the field, it turns out that creating pairs with rather large longitudinal
momentum is more likely to occur.

\begin{table}[!h]
\rule{0.48\textwidth}{0.5pt}
\caption{Characterization of resonances in a bOEF as shown in 
Figs.~\ref{fig:1dim-mom}(a) and \ref{fig:1dim-mom}(b) by the numbers of 
participating photons from the strong and weak modes. The 
corresponding plateau lengths to reach the maximum, quasienergies and 
momentum components are also indicated.}
\begin{center}
\begin{tabular}{|p{2.0cm}|p{1.6cm}|p{1.0cm}|p{1.0cm}|p{1.0cm}|p{1.0cm}|}
\hline
\parbox[0pt][1.8em][c]{0cm}{} Resonance & $n_1+n_2$ & $N_{\mathrm{max}}$ 
& $2 \bar\varepsilon_{\pmb{p}}\ [m]$ & $p_x\ [m]$ & $p_y\ [m]$\\
\hline
\hline
P00 & 4+1 & 16 & 2.4181 & 0.0 & 0.0\\
\hline
P10 & 9+0 & 213 & 2.6919 & 0.5716 & 0.0\\
\hline
P20 & 5+1, 1+2 & 104 & 2.7810 & 0.6702 & 0.0\\
\hline
P30 & 6+1, 2+2 & 158 & 3.0378 & 0.9049 & 0.0\\
\hline
P40 & 7+1, 3+2 & 332 & 3.3840 & 1.1707 & 0.0\\
\hline
P50 & 8+1, 4+2 & 145 & 3.6392 & 1.3485 & 0.0\\
\hline
P01 & 4+1, 0+2 & 56 & 2.4759 & 0.0 & 0.2740\\
\hline
P02 & 5+1, 1+2 & 30 & 2.7297 & 0.0 & 0.7069\\
\hline
P03 & 6+1, 2+2 & 52 & 3.0392 & 0.0 & 1.0114\\
\hline
P04 & 7+1 & 73 & 3.3397 & 0.0 & 1.2483\\
\hline
P05 & 3+2 & 97 & 3.3848 & 0.0 & 1.2809\\
\hline
P06 & 8+1 & 104 & 3.6410 & 0.0 & 1.4581\\
\hline
P07 & 4+2 & 96 & 3.6858 & 0.0 & 1.4878\\
\hline
\hline
\end{tabular}
\end{center}
\label{tab:photonNoVsMomenta}
\rule{0.48\textwidth}{0.5pt}
\end{table}

An additional important issue is related to the fact that photon numbers $n_2 > 1$ from the perturbative field mode can occur for some specific momenta (see also Sec.~III.A). 
Moreover it is worth mentioning that almost every determined resonance for nonzero momentum lies quite far below the maximum value of  $W(\pmb{p}=0,\tau)\approx 2$. This 
is mainly caused by an interaction time which is not sufficiently long to reach the highest possible $W(\pmb{p},\tau)$ for nonzero momenta; they would be reached for 
$N_{\mathrm{max}}>\Nplat = 16$ only (see the third column in Table III). Besides, the finite step size $\Delta p$ used in our computations may cause a situation in which the position of 
a resonance is not hit exactly. 

Finally, we want to emphasize that in both panels in Fig.~\ref{fig:1dim-mom} the area below $W(\pmb{p},\tau)$ resulting from a bOEF exceeds those corresponding to the cases 
driven by a monofrequent OEF (mOEF),  which are shown for comparison. These patterns already provide evidence that the density of created pairs can be substantially enhanced 
by superimposing a perturbative high-frequency mode onto a strong low-frequency mode.

\subsection{Two-dimensional momentum distribution}
\label{subsec:2d-mom}

\begin{figure}[!h]
\rule{0.48\textwidth}{0.5pt}\\
\subfigure[]{\includegraphics[width=0.52\textwidth]{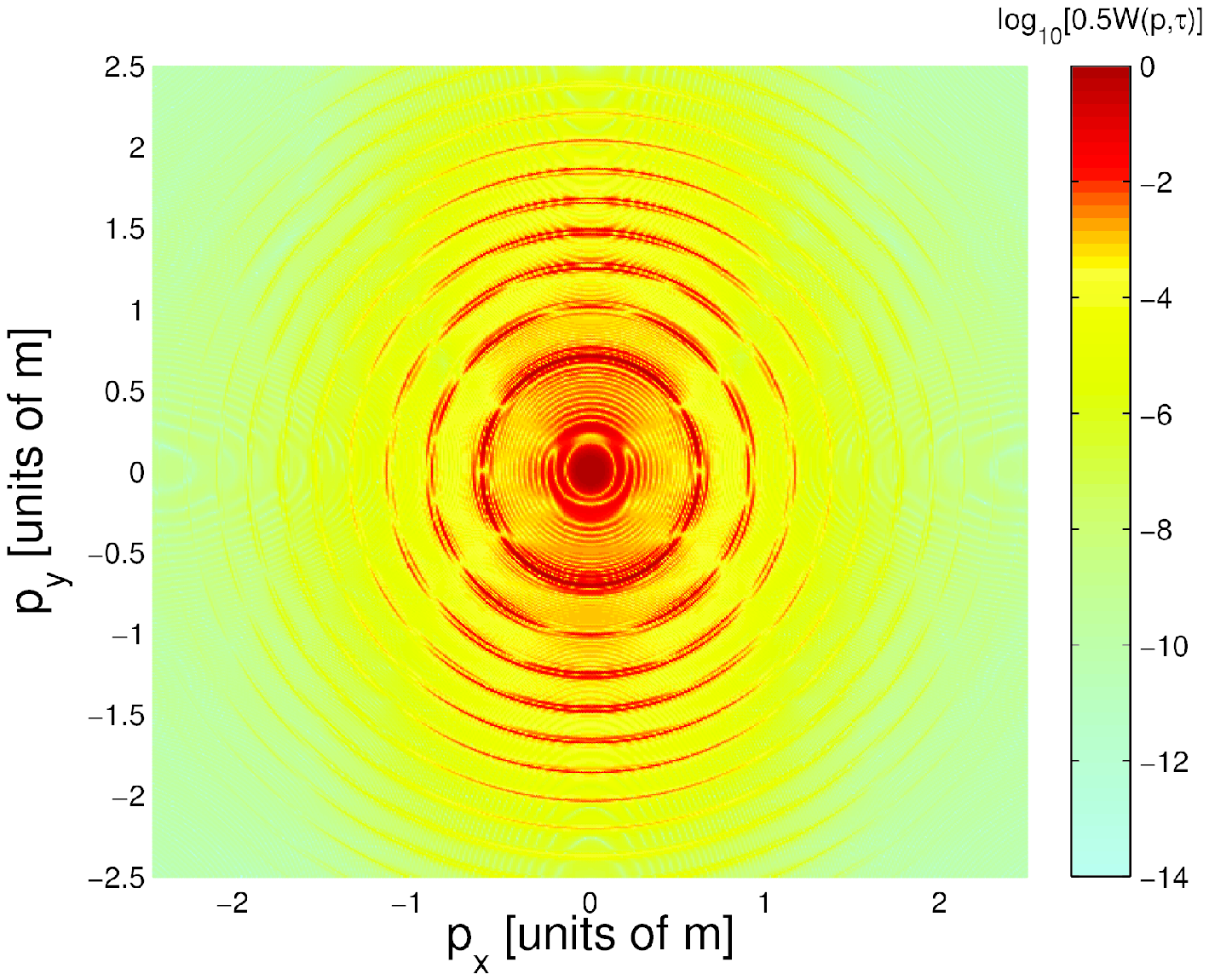}}\hfill\subfigure[]{\includegraphics[width=0.52\textwidth]{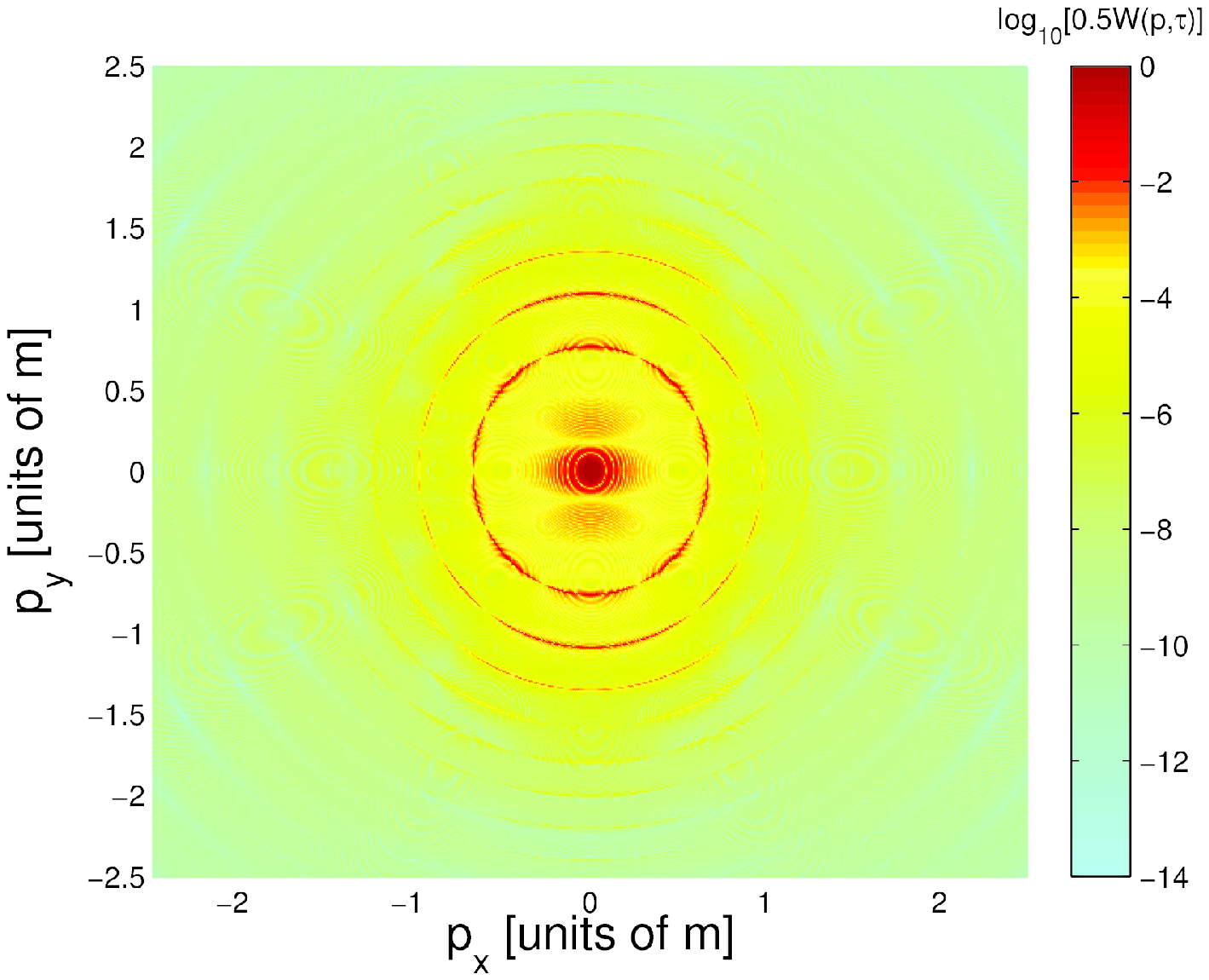}}
\caption{Two-dimensional momentum distributions in the $(p_{x},p_{y})$ plane for pair production in  (a) a bOEF [$\xi_1=1.0$, $\xi_2=0.1$, 
$\omega_1=0.3m$, $\omega_2=1.24385m$, and $\Nplat = 16$] and (b) mOEF [$\xi_1=1.0$, $\omega_1=0.34888 m$, and $\Nplat = 45$]. In the center of both 
panels a maximum is achieved. While in  (a) this maximum is characterized by  the {[4+1]}-process, in (b) it is despicted by  the {[7]}-process.
Observe that the single-particle distribution function is plotted in the form $\log_{10}[\frac{1}{2}W(\pmb{p},\tau)]$.}
\label{fig:2dim-mom}
\rule{0.48\textwidth}{0.5pt}
\end{figure}

While in the previous subsection the momentum dependence of the single-particle distribution function was studied for the longitudinal 
and transversal directions, now we shall examine in a more general way the two-dimensional momentum distribution in the $(p_x,p_y)$ plane. 
This way, all intermediate directions are covered. We vary both $p_x$ and $p_y$ in the interval from --2.5$m$ to 2.5$m$. The field parameters 
are chosen again as $\xi_1=1.0$, $\xi_2=0.1$, $\omega_1=0.3m$, $\omega_2=1.24385m$, and $\Nplat = 16$. They lead to a [4+1] resonance with 
maximal probability at the point $p_x = p_y = 0$. The corresponding distribution is shown in Fig.~\ref{fig:2dim-mom}(a) in a color-coded 
scheme. The logarithmic quantity $\log_{10}[\frac{1}{2}W(\pmb{p},\tau)]$ has been plotted.  We observe ring-shaped regions of high probability 
which reflect the resonance properties of the process. Cuts along the $p_x$ ($p_y$) axis at $p_y=0$ ($p_x=0$) would reproduce the one-dimensional 
momentum distributions along the transversal  (longitudinal) directions shown in Fig.~\ref{fig:1dim-mom}. This way, the resonance rings can 
be characterized in terms of the numbers $n_1$ and $n_2$ of participating photons. We note that the ring-shaped structures are slightly 
elongated in the longitudinal direction. This is because large values of $p_y$ are more likely to occur than large values of $p_x$, as was 
mentioned before.

It is interesting to compare the two-dimensional momentum distribution in a bOEF with that in a mOEF. To this end, we use the following mOEF 
parameters~$\xi_1 = 1.0$, $\omega_1 = 0.34888 m$ and $\Nplat = 45$, which also lead to a resonance with maximum probability at $p_x = p_y = 0$. 
It corresponds to the absorption of $n=7$ photons (see Table I in Ref.~\cite{mocken2010}). Figure~\ref{fig:2dim-mom}(b) presents the corresponding 
two-dimensional distribution function on the same color-coded logarithmic scale. We  observe similar ring-shaped structures, with the expected [7] 
resonance in the center. This one  is surrounded by further less pronounced resonances for increasing photon numbers $n=8, 9, 10, \ldots$,  
a fact which resembles some  finding of Ref.~\cite{Blaschke2013}.  They 
fulfill the monofrequent resonance condition $2 \bar\varepsilon_{\pmb{p}} \approx n \omega_1$. A rather remarkable trait is the distance between 
the resonance rings. By comparing Figs.~\ref{fig:2dim-mom}(a) and \ref{fig:2dim-mom}(b) it becomes evident that for the bOEF, these are more closely spaced than 
for the mOEF. The reason is that the presence of a second frequency in the case of a bOEF offers more options to fulfill the resonance condition 
in Eq.~\eqref{eqn:resonance-condition} than in the monofrequent case. One can observe, moreover, that the number of resonance rings with sizeable 
magnitudes is increased in the presence of a bOEF. This indicates that the total pair yield, which is obtained by integration over the momentum 
space, will be substantially larger in the bOEF. This becomes especially apparent in view of the considerably larger red-colored regions in 
Fig.~\ref{fig:2dim-mom}(a).  
 
\subsection{Total number of electron-positron pairs}
\label{subsec:total-numbers}

From the two-dimensional momentum distributions in the $(p_x,p_y)$ plane, we can obtain the total number of produced electron-positron pairs per unit 
volume by evaluating the integral \eqref{eqn:tot-no}. We can exploit the cylindrical symmetry of the problem about the $y$ axis and, accordingly, perform the integral over 
$\dbar^3 p$ in cylindrical coordinates using $|p_x|$ as the polar coordinate. For a finite interaction time $\tau$, we thus arrive at the expression
\begin{eqnarray}
 \mathpzc{N}_{\ e^-e^+} = \frac{1}{4 \pi^2} \int_{-\infty}^\infty \int_0^\infty W(p_x,p_y,0,\tau) |p_x| dp_x dp_y .
\end{eqnarray}
For the numerical evaluation, it is required to introduce a finite cutoff for such values in the momentum plane where the distribution function no  longer contributes 
appreciably. We shall use the cutoff value $\bar{p} = 3 m$ and restrict the ranges of integration to $|p_x|\le \bar{p}$ and 
$|p_y|\le\bar{p}$. Hence,  we obtain approximately 
\begin{eqnarray}
 \mathpzc{N}_{\ e^-e^+} \approx \frac{1}{8 \pi^2} \int_{-\bar p}^{+\bar p} \int_{-\bar p}^{+\bar p} W(p_x,p_y,0,\tau) |p_x| dp_x dp_y,
\end{eqnarray}
where we have exploited the symmetry relation $W(-p_x,p_y,0,\tau) = W(+p_x,p_y,0,\tau)$ in order to make the connection of the total number of produced 
pairs with the two-dimensional momentum distributions, as illustrated in Fig.~\ref{fig:2dim-mom}, explicit. The latter integral is solved by numerical 
integration as a sum of the corresponding discretized values provided by the two-dimensional data sets. 

\begin{figure}[!h]
\rule{0.48\textwidth}{0.5pt}\\
\hspace*{-1em}\subfigure[\ Dependence on the strong-field intensity parameter for bOEF versus mOEF.]
{\includegraphics[width=0.48\textwidth]{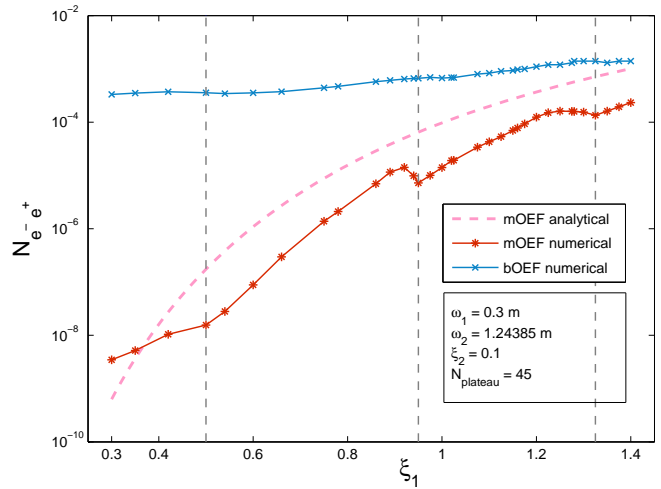}}\hfill
\subfigure[\ Dependence on the combined Keldysh parameter for various bOEFs.]
{\includegraphics[width=0.48\textwidth]{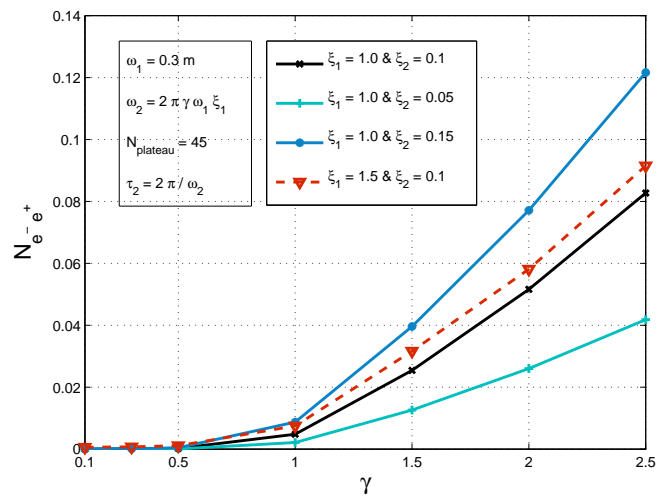}}
\caption{Total number of electron-positron pairs produced in a unit volume $V_C$. Panel (a) shows the dependence on the strong-field intensity parameter $\xi_1$ 
for a bOEF with parameters as indicated in the legend (blue crosses). For comparison, corresponding numerical results for a mOEF (obtained by setting $\xi_2=0$) 
are shown (red asterisks), as well as the analytical prediction from Ref.~\cite{Brezin} (pink dashed line). The vertical grey dashed lines mark positions where channel 
closings for the mOEF appear. Panel (b) shows the dependence on the combined Keldysh parameter $\gamma = m / (e E_1 \tau_2)$.}
\label{fig:totN}
\rule{0.48\textwidth}{0.5pt}
\end{figure}

First, we analyze the dependence of the total number of produced pairs on the strong-field intensity parameter of a bOEF. The latter has been varied within 
the interval $0.3\le\xi_1\le 1.4$. The other parameters were set to $\omega_1 = 0.3 m$, $\omega_2 = 1.24385 m$, $\xi_2 = 0.1$ and $\Nplat = 45$. The number 
of plateau cycles has been chosen to enable a good comparison with the mOEF case.\footnote{Note that, since $N_{\mathrm{max}}=16$ for the bOEF according to 
Table~I, the  pair production probability will reach its next maximum close to $\Nplat = 45$, due to the periodic Rabi oscillations.} The latter was realized 
by using parameters identical to the strong mode of the bOEF. The respective plot for the total number of produced electron-positron pairs per unit volume 
(given by the Compton volume $V_C = m^{-3}$) is shown in Fig.~\ref{fig:totN}(a). 

We observe that throughout the whole range of $\xi_1$ values, the pair yield from the bOEF is substantially larger than that from the mOEF. This indicates an 
effective enhancement of the pair production by superimposing a second weak mode of high frequency in addition to a stronger one. In particular, when 
$\xi_1 \ll 1$, the bOEF results exceed the mOEF results by $5$ orders of magnitude. The (average) slope of the mOEF curve, however, is larger than the bOEF 
curve (see also Refs.~\cite{dgs2008,DiPiazza:2009py,Jansen2013,Augustin2014}), so that eventually for $\xi_1\gg 1$ the enhancing effect of the perturbative mode 
will diminish. This behavior coincides with the intuitive expectation according to which the perturbative high-frequency mode is getting less relevant because 
the low-frequency mode becomes strong enough to dominate the production process. 

Another noticeable issue shown in Fig.~\ref{fig:totN}(a) is the comparison for the mOEF case between the numerical results and the analytical curve obtained 
from Ref.~\cite{Brezin}. The general trend of both curves is very similar, with the caveat that their absolute heights are somewhat different (see also Fig.~14 in 
Ref.~\cite{mocken2010}).
A remarkable feature of the numerical curve is that it exhibits considerable dropoffs around the values of $\xi_1 \approx 0.5$, 0.95, and 1.32, which do not 
occur for the analytical curve. They can be attributed to the fact that at these specific $\xi_1$ values a resonance disappears due to energetic reasons. Namely, 
when $\xi_1$ increases, the effective mass $m_*$ increases as well, which determines the minimal energy to be absorbed from the field in order to produce a pair. 
This phenomenon is commonly known as a channel-closing effect (see, e.g., Ref.~\cite{HReiss,Kohlfurst:2013ura}). The positions of the channel closings are marked by the vertical dashed 
lines in Fig.~\ref{fig:totN}(a). From left to right, they correspond to the vanishing of the resonances with $n=7$, 8, and 9, respectively.

Finally, in addition to the previous discussion, Fig.~\ref{fig:totN}(b)  shows the dependency of the total number of produced pairs per unit  volume $V_C$ on the combined Keldysh parameter given by
$\gamma = \frac{m}{eE_1 \tau_2}$, where $\tau_2 = 2 \pi / \omega_2$ is the period of the perturbative mode. The interval $0.1\le\gamma\le2.5$ is considered and the Keldysh parameter is varied 
by varying $\omega_2$ while keeping 
$E_1$ fixed. The curve for $\xi_1=1.0$, $\xi_2=0.1$ (black solid line) may serve us as a reference. By comparison we find that, when the intensity parameter of the weak mode is amplified by a 
factor of 1.5 (blue solid line), a much stronger growth of the particle number results as compared with the case when the intensity parameter of the strong mode is increased by the same factor 
(red dashed line). Thus, in the considered interaction regime, it is more efficient to invest in an increase of the small parameter $\xi_2$ than a further increase of the large parameter $\xi_1$.

Moreover, a comparison of the solid lines in Fig.~\ref{fig:totN}(b) reveals that the particle number scales linearly with $\xi_2$. From our formalism, this can be understood by inspection of 
Eq.~(\ref{vlasov}). The  total electric field $E(t)=E_1(t)+E_2(t)$ enters quadratically there, and since we operate in a regime where the combined 
effect of both modes is crucial, one may expect that the resulting cross term $\sim E_1(t)E_2(t)$ gives the largest contribution to the single-particle distribution function. This leads to 
a linear dependence on $E_2$, in accordance with the numerical outcomes.

We would like to stress that, despite its very plausible explanation, the linear scaling with the parameter $\xi_2$ is an interesting feature of PP in a bOEF. In fact, other PP processes, which 
exhibit a similar enhancement mechanism in a bifrequent field, do not share this scaling property. For dynamically assisted Breit-Wheeler~\cite{Jansen2013} as well as Bethe-Heitler 
\cite{DiPiazza:2009py,Augustin2014} PP in strong laser fields, a quadratic scaling law $\sim\xi_2^2$ has been found, provided that a single high-frequency photon assists in the process. This is 
because the underlying PP amplitude there  is linear in $\xi_2$ whose square provides the PP rate. That a subquadratic dependence on $\xi_2$ arises in a bOEF can already be anticipated on the 
basis of its resonant nature, which does not exist for the PP processes in Refs.~\cite{DiPiazza:2009py,Augustin2014,Jansen2013}. The resonances provide a substantial contribution to the total number 
of produced pairs, but their height is bounded by the maximum value of  $W({\pmb{p}},\tau)=2$. Thus, an increase of $\xi_2$ cannot enhance this value further. However, the fundamental reason for 
the different $\xi_2$ scaling in a bOEF as compared with intense bifrequent laser fields~ \cite{DiPiazza:2009py,Augustin2014,Jansen2013} can be traced back to the fact that the strong mode of the 
bOEF alone is capable of producing pairs, whereas the quantum vacuum is stable in the presence of a plane-wave laser field. Therefore, in the latter case,  the leading-order vacuum polarization 
diagram, whose imaginary part is related to the PP rate via the optical theorem, contains the perturbatively weak field at two vertices. In contrast, to the  PP in an OEF also diagrams with just a 
single perturbative vertex contribute,  which leads to the observed linear dependence on the small parameter $\xi_2$.

\section{Conclusion}

We presented a comprehensive investigation of electron-positron pair production in a bifrequent OEF composed of a strong low-frequency and a perturbative high-frequency mode. 
A quantum kinetic approach allowed us to calculate the single-particle distribution function by using numerical methods. We found a pronounced resonant behavior depending on 
the frequency composition of the field and demonstrated Rabi-like oscillations in full time resolution on and slightly off a resonance peak (Secs.~\ref{subsec:res-spec}~and~\ref{subsec:rabi-like}). 
These numerical findings showed a clear agreement with predictions from the theory [Sec.~\ref{subsec:res-effects}]. The occurrence of the resonances 
could be explained by discrete numbers of photons absorbed from (or emitted into) the bifrequent field. While in the case of vanishing particle momenta, only odd total photon 
numbers are allowed due to charge-conjugation symmetry, resonant production of pairs with nonzero momenta may also proceed involving an even number (Secs.~\ref{subsec:1d-mom}~and~\ref{subsec:2d-mom}). 
Besides, it has been shown that  processes with more than one photon from the perturbative mode also occur. Finally, we have seen that the superimposed 
perturbative mode can substantially amplify the total number of generated electron-positron pairs in comparison with the case where a single strong mode drives the vacuum decay 
(Sec.~\ref{subsec:total-numbers}). The enhancement shows a linear dependence on the intensity parameter of the weak mode and is especially pronounced at large values of the 
relative Keldysh parameter (i.e., at relatively low field strengths of the strong mode and high frequencies of the weak mode).

\section*{Acknowledgments}

S. Villalba-Ch\'avez  gratefully acknowledges the support of  the Alexander von Humboldt Foundation in the early stage of this project.

\appendix

\section{Remarks on the second  quantization \label{subsec:gen-remarks}}

This appendix provides  some details about the second quantization formalism in a time-dependent electric field of the form $\pmb{E}(t)=(0,E(t),0)$. To begin with,   
we remark  that  the  action of the creation  operators  $a_{\pmb{p},\mathpzc{s}}^\dagger(t)$  and  $b_{-\pmb{p},\mathpzc{s}}^\dagger(t)$ on the instantaneous vacuum state 
$\vert\mathrm{VAC},t\rangle$ of the theory, defined by 
\begin{eqnarray}
a_{\pmb{p},\mathpzc{s}}(t)\vert\mathrm{VAC},t\rangle=b_{-\pmb{p},\mathpzc{s}}(t)\vert\mathrm{VAC}, t\rangle =0\ ,
\end{eqnarray} allows us to  build  up a Fock space for the quasiparticles.  A connection between  the previous set of states   and  those resulting  at 
$t\to t_{\rm in}$ can   be established  by means of  a canonical unitary operator $\mathpzc{U}(t,t_{\mathrm{in}})$ so that 
\begin{eqnarray}\label{vacummsrelation}
\vert\mathrm{VAC},t\rangle=\mathpzc{U}(t,t_{\mathrm{in}})\vert\mathrm{VAC},\mathrm{in}\rangle,
\end{eqnarray}where the ground state $\vert\mathrm{VAC},\mathrm{in}\rangle$  is defined in the Heisenberg picture by $a_{\mathrm{in}}\vert\mathrm{VAC},\mathrm{in}\rangle=b_{\mathrm{in}}\vert\mathrm{VAC}, \mathrm{in}\rangle =0$ 
with  $a_{\mathrm{in}}\equiv a_{\pmb{p},\mathpzc{s}}$ and $b_{\mathrm{in}}\equiv b_{-\pmb{p},\mathpzc{s}}$ at $t\to t_{\rm in}$. With these details in mind 
and by considering  Eq.~(\ref{vacummsrelation}),  the following transformation properties are obtained:
\begin{eqnarray}\label{unitaryBogolyubovoperator}
&&a_{\pmb{p},\mathpzc{s}}(t)=\mathpzc{U}(t,t_{\mathrm{in}}) a_\mathrm{in} \mathpzc{U}^\dagger(t,t_{\mathrm{in}}),\\ &&b_{-\pmb{p},\mathpzc{s}}^\dagger(t)=\mathpzc{U}^\dagger(t,t_{\mathrm{in}}) b_{\mathrm{in}}^\dagger \mathpzc{U}(t,t_{\mathrm{in}}).
\end{eqnarray} 

The  structure of the   evolution operator $\mathpzc{U}(t,t_{\mathrm{in}})$  has been investigated in Refs.~\cite{Bagrov,Gitman,Fradkin}. Its  construction  
can be carried out by following a procedure similar to  that established  in the context of  the  BCS theory (for further details,  we refer the reader 
to Ref.~\cite{Blaizot} and references therein). Following the ansatz  of these references, we  express  $\mathpzc{U}(t,t_{\mathrm{in}})$ in the following  form:
\begin{eqnarray}\label{unitaryopertaor}
\begin{array}{c}
\displaystyle \mathpzc{U}(t,t_{\mathrm{in}})=\exp[\Lambda(t,t_{\mathrm{in}})],\quad \Lambda(t,t_{\mathrm{in}})=\sum_{\pmb{p},\mathpzc{s}} \Lambda_{\pmb{p},\mathpzc{s}}(t,t_{\mathrm{in}}),\\ 
\Lambda_{\pmb{p},\mathpzc{s}}(t,t_{\mathrm{in}})=\alpha a_{\mathrm{in}}^\dagger b_{\mathrm{in}}^\dagger-\alpha^* b_{\mathrm{in}}a_{\mathrm{in}}+i\beta a_{\mathrm{in}}^\dagger a_{\mathrm{in}}-i\beta b_{\mathrm{in}} b_{\mathrm{in}}^\dagger
\end{array}
\end{eqnarray}where the respective complex  and real parameters $\alpha$ and $\beta$ are  functions of $\pmb{p}$ and $t$. It is worth mentioning at this point that  
the above operator $\mathpzc{U}(t,t_{\mathrm{in}})$ preserves both  the  unitary and the canonic features. To  establish  the relation  between $\alpha$   and the 
coefficient arising from the  Bogolyubov transformations,  we  Taylor-expand  all instances of $\mathpzc{U}(t,t_{\mathrm{in}})$  in  Eq.~(\ref{unitaryBogolyubovoperator}) so that 
\begin{eqnarray}\label{unitaryBogolyubovoperator1}
\begin{array}{c}
\displaystyle a_{\pmb{p},\mathpzc{s}}(t)=\sum_{n=0}^{\infty}\frac{1}{n!} \left[\Lambda,\left[\Lambda,\ldots\left[\Lambda,a_\mathrm{in}\right]\ldots\right]\right],\\
\displaystyle b_{-\pmb{p},\mathpzc{s}}^\dagger(t)=\sum_{n=0}^{\infty}\frac{1}{n!} \left[\Lambda,\left[\Lambda,\ldots\left[\Lambda,b_\mathrm{in}^\dagger\right]\ldots\right]\right].
\end{array}
\end{eqnarray}As a consequence of a reiterated use of the  commutation rules $
\left[\Lambda, a_{\mathrm{in}}\right]=-\alpha b_{\mathrm{in}}^\dagger-i\beta a_{\mathrm{in}}$,   $\left[\Lambda, b_{\mathrm{in}}\right]=\alpha a_{\mathrm{in}}^\dagger-i\beta b_{\mathrm{in}}$, 
$[\Lambda, a_{\mathrm{in}}^\dagger]=-\alpha^* b_{\mathrm{in}}+i\beta a_{\mathrm{in}}^\dagger$ and $[\Lambda, b_{\mathrm{in}}^\dagger]=\alpha^*a_{\mathrm{in}}+i\beta b_{\mathrm{in}}^\dagger$,  
we can reduce  the above relations to 
\begin{eqnarray}
\left[\begin{array}{c} 
a_{\pmb{p},\mathpzc{s}}(t)\\ b_{-\pmb{p},\mathpzc{s}}^\dagger(t) 
\end{array}\right]=
\left[\begin{array}{cc}
g^*(\pmb{p},t) & f(\pmb{p},t) \\ -f^*(\pmb{p},t)  &  g(\pmb{p},t)       
\end{array}
\right]
\left[\begin{array}{c}
a_{\mathrm{in}}\\ b_{\mathrm{in}}^\dagger
\end{array}\right],
\end{eqnarray} where the  unknown matrix elements 
\begin{eqnarray}
&&f(\pmb{p},t)=-\frac{\alpha}{\sqrt{\vert\alpha\vert^2+\beta^2}}\sin(\sqrt{\vert\alpha\vert^2+\beta^2}),\\
&&g(\pmb{p},t)=\cos(\sqrt{\vert\alpha\vert^2+\beta^2})+i\frac{\beta}{\alpha}f(\pmb{p},t)
\end{eqnarray} clearly satisfy  the condition $\vert g(\pmb{p},t)\vert^2+\vert f(\pmb{p},t)\vert^2=1$. 

Let us turn our attention to the connection between the $\mathrm{in}-$ and the  instantaneous ground states [Eq.~(\ref{vacummsrelation})].   
When $\mathpzc{U}(t,t_{\mathrm{in}})$ is Taylor-expanded, its  action  on $\vert\mathrm{VAC},\mathrm{in}\rangle$ involves a rather complicated mixture 
of the second quantization operators $a_{\mathrm{in}},\ b_{\mathrm{in}},\ a_{\mathrm{in}}^\dagger,\ b_{\mathrm{in}}^\dagger$. 
To avoid this problem, we disentangle the evolution operator according to the  rule  $\mathpzc{U}(t,t_{\mathrm{in}})=\prod_{\pmb{p},\mathpzc{s}}\exp[\Lambda_{\pmb{p},\mathpzc{s}}(t,t_{\mathrm{in}})]$.
This step is allowed due to the fact that the commutator between two arbitrary elements $\Lambda_{\pmb{p},\mathpzc{s}}(t,t_{\mathrm{in}})$ and $\Lambda_{\pmb{p}^\prime,\mathpzc{s}^\prime}(t,t_{\mathrm{in}})$ 
vanishes identically. Next, we  expand $\exp[\Lambda_{\pmb{p},\mathpzc{s}}(t,t_{\mathrm{in}})]$ and use  the   identities  
$\Lambda_{\pmb{p},\sigma}^{2n+1}(t,t_{\mathrm{in}})=(-1)^n(\vert\alpha\vert^{2}+\beta^2)^n\Lambda_{\pmb{p},\sigma}(t,t_{\mathrm{in}})$ and $\Lambda_{\pmb{p},\sigma}^{2n}(t,t_{\mathrm{in}})=(-1)^{n-1}(\vert\alpha\vert^2+\beta^2)^{n-1}\Lambda_{\pmb{p},\sigma}^2(t,t_{\mathrm{in}})$
with $\Lambda_{\pmb{p},\sigma}^2(t,t_{\mathrm{in}})=-(\vert\alpha\vert^2+\beta^2)(b_{\mathrm{in}}^\dagger b_{\mathrm{in}} a_{\mathrm{in}}^\dagger a_{\mathrm{in}}+b_{\mathrm{in}} b_{\mathrm{in}}^\dagger a_{\mathrm{in}} a_{\mathrm{in}}^\dagger)$ 
to  express the instantaneous ground state of the theory as a two-mode squeezed state of the $\mathrm{in}-$particle pairs
\begin{equation}
\vert\mathrm{VAC}, t\rangle=\prod_{\pmb{p},\mathpzc{s}}g^*(\pmb{p},t)\exp\left[\frac{f(\pmb{p},t)}{g^*(\pmb{p},t)}b_{\mathrm{in}}^\dagger a_{\mathrm{in}}^\dagger\right]
\vert\mathrm{VAC}, \mathrm{in}\rangle.
\end{equation} Considering this expression, it is quite simple to verify  that the vacuum persistence probability  is given by
\begin{eqnarray}\label{persistenceprobability}
 \mathpzc{P}_{\mathrm{vac}}(t)&=&\vert\langle\mathrm{VAC}, t\vert\mathrm{VAC}, \mathrm{in}\rangle\vert^2=\exp\sum_{\pmb{p},\mathpzc{s}}\ln\left[g(\pmb{p},t)^2\right]\nonumber\\
 &=&\exp\left[(t-t_{\mathrm{in}}) V \varGamma_{\mathrm{vac}}(t)\right],
\end{eqnarray}where  $\varGamma_{\rm vac}(t)$ denotes the  instantaneous rate of vacuum decay per unit of volume:
\begin{eqnarray}\label{vacdecrate}
&&\varGamma_{\rm vac}(t)=\frac{\ln[\mathpzc{P}_{\mathrm{vac}}(t)]}{(t-t_{\mathrm{in}})V}\nonumber\\&&\qquad\quad=\frac{2}{t-t_{\mathrm{in}}}\int \dbar^3 p\ln\left(1-\vert f(\pmb{p},t)\vert^2\right).
\end{eqnarray}This last expression results from  the  transition to the infinite volume continuum limit in which the relation 
$\frac{1}{V}\sum_{\pmb{p}}\to\int \dbar^3p$ with $\dbar^3\equiv d^3/(2\pi)^3$ holds. The factor of $2$ in  Eq.~(\ref{vacdecrate}) 
results from the sum over the discrete spin variable $\mathpzc{s}$. In the subcritical regime [$E\ll E_c$], the use of the above expression  
allows us to reproduce the vacuum decay rate $\varGamma_{\mathrm{vac}}=\frac{(eE)^2}{4\pi^3}\sum_{n=1}^\infty\frac{1}{n^2}\exp[-\pi\frac{E_c}{E}]$
found  by Schwinger for  a constant electric field. A detailed calculation of 
this subject can be found in  \cite{Tanji:2008ku}. 

\section{Deriving the   Boltzmann-Vlasov equation \label{appb}}

The integrodifferential representation  of the quantum kinetic  equation for the \pp process [Eq.~(\ref{vlasov})] is determined only  
after performing some additional steps in Eqs.~(\ref{firstequa}) and (\ref{secondequa}).  To this end, we determine  the temporal  
equations of  new  variables $\bar{f}(\pmb{p},t)$ and $\bar{g}(\pmb{p},t)$, which are linked  to  the original coefficients $f(\pmb{p},t)$ 
and $g(\pmb{p},t)$ through unitary transformations  $\bar{f}(\pmb{p},t)=-if(\pmb{p},t)\exp\left[i\int_{t_0}^t dt^\prime \mathpzc{a}_{\pmb{p}}(t^\prime)\right]$ 
and $\bar{g}(\pmb{p},t)=ig(\pmb{p},t)\exp\left[-i\int_{t_0}^t dt^\prime \mathpzc{a}_{\pmb{p}}(t^\prime)\right]$. These details, 
together with Eqs.~(\ref{firstequa})~and~(\ref{secondequa}), allow us to establish  the system of \odes given by Eq.~(\ref{intermdiate1v}) and Eq.~(\ref{intermdiate2v}):
\begin{eqnarray}\label{intermdiate1}
&\displaystyle\dot{\bar{f}}(\pmb{p},t)=-\frac{1}{2}Q_{\pmb{p}}(t)\bar{g}(\pmb{p},t)\exp\left[2i\Theta_{\pmb{p}}(t)\right],\\ &\displaystyle\dot{\bar{g}}(\pmb{p},t)=\frac{1}{2}Q_{\pmb{p}}(t)\bar{f}(\pmb{p},t)\exp\left[-2i\Theta_{\pmb{p}}(t)\right]\label{intermdiate2}
\end{eqnarray}with $Q_{\pmb{p}}(t)$ and $\Theta_{\pmb{p}}(t)$ being defined as 
\begin{equation}\label{parameters}
Q_{\pmb{p}}(t)=\frac{eE(t)\epsilon_\perp}{\mathpzc{w}_{\pmb{p}}^2(t)},\quad \Theta_{\pmb{p}}(t)=\int_{t_0}^t dt^\prime\ \mathpzc{w}_{\pmb{p}}(t^\prime).
\end{equation} 
In this framework, the initial conditions $\bar{f}(\pmb{p},-\infty)=0$ and $\bar{g}(\pmb{p},-\infty)=1$ ensure the pure 
vacuum condition. At this point it is convenient to look at the time evolution equations of the single-particle distribution function Eq.~(\ref{spdf})  and  the  
quasiparticle correlation function $O(\pmb{p},t)=\sum_{\mathpzc{s}}\langle\mathrm{VAC}, \mathrm{in}\vert b_{-\pmb{p},\mathpzc{s}}^\dagger(t)a_{\pmb{p},\mathpzc{s}}^\dagger(t)\vert\mathrm{VAC}, \mathrm{in}\rangle=2f(\pmb{p},t)g^*(\pmb{p},t)$.  
With the help of Eqs.~(\ref{intermdiate1}) and (\ref{intermdiate2}), we find that
\begin{eqnarray}
\label{kinI} &&\dot{W}(\pmb{p},t)=-Q_{\pmb{p}}(t)\mathrm{Re}\left\{O(\pmb{p},t)\exp\left[-2i\Theta_{\pmb{p}}(t)\right]\right\},\\
\label{kinI2} &&\dot{O}(\pmb{p},t)=-Q_{\pmb{p}}(t)\left[1-W(\pmb{p},t)\right]\exp\left[2i\Theta_{\pmb{p}}(t)\right].
\end{eqnarray}
Next, we integrate the second expression in Eq.~\eqref{kinI2} over time  from $-\infty$ to $t$ and insert the resulting expression 
for $O(\pmb{p},t)$  into Eq.~(\ref{kinI}) afterwards.  The outcome of this procedure is the   Boltzmann-Vlasov  equation [Eq.~(\ref{vlasov})].

\end{document}